\documentclass[12pt]{article}
\usepackage{amsmath,epsf,amssymb,latexsym,cite,xy}
\xyoption{matrix}\xyoption{arrow}

\usepackage[dvips]{epsfig}

\newcommand{\beq}{\begin{equation}}
\newcommand{\eeq}{\end{equation}}
\newcommand{\beqa}{\begin{eqnarray}}
\newcommand{\eeqa}{\end{eqnarray}}

\newif\iffigs\figstrue

\DeclareFontFamily{U}{rsf}{}
\DeclareFontShape{U}{rsf}{m}{n}{
   <5> <6> rsfs5 <7> <8> <9> rsfs7 <10-> rsfs10}{}
\DeclareMathAlphabet\Scr{U}{rsf}{m}{n}

\def\pplogo{\vbox{\kern-\headheight\kern -43pt
\halign{##&##\hfil\cr&{
\ppnumber}\cr\rule{0pt}{2.5ex}&\ppdate\cr}
}}
\makeatletter
\def\ps@firstpage{\ps@empty \def\@oddhead{\hss\pplogo}%
   \let\@evenhead\@oddhead 
}
\def\maketitle{\par
  \begingroup
  \def\thefootnote{\fnsymbol{footnote}}
  \def\@makefnmark{\hbox{$^{\@thefnmark}$\hss}}
  \if@twocolumn
  \twocolumn[\@maketitle]
  \else \newpage
  \global\@topnum\z@ \@maketitle \fi\thispagestyle{firstpage}\@thanks
  \endgroup
  \setcounter{footnote}{0}
  \let\maketitle\relax
  \let\@maketitle\relax
  \gdef\@thanks{}\gdef\@author{}\gdef\@title{}\let\thanks\relax}
\makeatother

\def\la{~\mbox{\raisebox{-.6ex}{$\stackrel{<}{\sim}$}}~}

\def\O{\Scr{O}}

\def\R{{\mathbb R}}
\def\Z{{\mathbb Z}}

\def\tr{\operatorname{tr}}

\def\cR{{\Scr R}}

\def\cL{{\Scr L}}

\def\cO{{\Scr O}}
\def\cV{{\Scr V}}

\def\LARGE{\large\bf}

\def\vev#1{{\langle #1 \rangle}}
\def\p{\partial}
\def\d{\partial}

\def\half{\frac {1}{2}}

\def\tV{{\tilde V}}

\def\mpl{m_{pl}}
\def\tphi{{\tilde\phi}}

\def\be{\begin{equation}}
\def\ee{\end{equation}}
\def\ba{\begin{eqnarray}}
\def\ea{\end{eqnarray}}

\begin{document}

\setcounter{page}0

\def\ppnumber{\vbox{\baselineskip14pt
\hbox{BRX-TH 622}
\hbox{hep-th/yymmnnn}}}
\def\ppdate{December 2010} \date{}

\title{\LARGE An Ignoble Approach to Large Field Inflation\\[10mm]}
\author{
Nemanja Kaloper${}^{1}$, Albion Lawrence${}^{2,3}$, and Lorenzo Sorbo${}^4$ \\[2mm]
\normalsize ${}^1$ Department of Physics, University of California, Davis, CA 95616 \\
\normalsize ${}^2$ Theory Group, Martin Fisher School of Physics, Brandeis University, \\
\normalsize MS057, PO Box 549110, Waltham, MA 02454\\
\normalsize${}^3$ CCPP, Department of Physics, New York University, NY, NY, 10003\\
\normalsize ${}^4$ Department of Physics, University of Massachusetts, Amherst, MA 01003\\
}

{\hfuzz=10cm\maketitle}

\begin{abstract}
We study an inflationary model developed by Kaloper and Sorbo, in which the inflaton is an axion with a sub-Planckian decay constant, whose potential is generated by mixing with a topological $4$-form field strength. This gives a 4d construction of ``axion monodromy inflation": the axion winds many times over the course of inflation and draws energy from the 4-form.  The classical theory is equivalent to chaotic inflation with a quadratic inflaton potential.  Such models can produce ``high scale" inflation driven by energy densities of the order of $(10^{16}\ GeV)^4$, which produces primordial gravitational waves potentially accessible to CMB polarization experiments.
We analyze the possible corrections to this scenario from the standpoint of 4d effective field theory, identifying the physics which potentially suppresses dangerous corrections to the slow-roll potential.  This yields a constraint relation between the axion decay constant, the inflaton mass, and the 4-form charge. We show how these models can evade the fundamental constraints which typically make high-scale inflation difficult to realize.  Specifically, the moduli coupling to the axion-four-form sector must have masses higher than the inflationary Hubble scale ($\la\ 10^{14}\ GeV$). There are also constraints from states that become light due to multiple windings of the axion, as happens in explicit string theory constructions of this scenario. Further, such models generally have a quantum-mechanical ``tunneling mode" in which the axion jumps between windings, which must be suppressed. Finally, we outline possible observational signatures. 
\end{abstract}

\vfil\break

\setlength{\topmargin}{-1cm}

\section{Introduction}

Simple models of inflation -- for example, a quadratic or quartic potential for a scalar field coupled to gravity -- work well as quantum theories.  To be sure, one must tune the model to maintain slow roll over sufficient efoldings and produce the observed density fluctuations; for a potential of the form
\be\label{eq:polynpot} V = m_{pl}^{4-n} g_n \phi^n \, ,
\ee
$g_2$ must be of order $10^{-12}$ and $g_4 \sim 10^{-14}$.\footnote{We will use the reduced Planck mass $m_{pl} \sim 2 \times 10^{18}\ GeV$ in this paper.}  But after this initial choice of couplings, 
this scenario is stable to quantum corrections from loops of inflatons and gravitons below the cutoff,\footnote{We will review this further in \S3, and thank N. Arkani-Hamed for discussions of this issue.} at least in the regime that produces fluctuations inside our current horizon. These corrections are controlled by the very softly broken shift symmetry that slow roll demands in the first place. The $n=2$ model in particular provides an excellent fit to the current WMAP data.

The problem with these effective field theories lies in justifying them via a sensible UV completion. Models of the form (\ref{eq:polynpot}) require super-Planckian expectation values to generate sufficient inflation to solve the flatness problem. More generally, such field ranges are required by any inflationary model producing observable tensor modes \cite{Lyth:1996im,Efstathiou:2005tq}, which includes the models (\ref{eq:polynpot}).  In this context, even highly irrelevant $m_{pl}$-suppressed corrections may be dangerous -- any corrections of the form $\phi^n/m_{pl}^{n-4}$ with order 1 coefficients may be sufficient to spoil the slow-roll conditions when $\phi > m_{pl}$.  

Since gravity is nonrenormalizable, unitarity will require Planck-suppressed operators.  One might think that a softly broken shift symmetry at tree level would help suppress the dangerous ones.  But Hawking radiation, gravitational instantons, and wormholes all break continuous global symmetries \cite{Kallosh:1995hi}, and continuous symmetries in string theory are generally either gauged or broken. Thus, there is no {\it a priori} reason for the coefficients of any symmetry-violating Planck-suppressed operators to be small.  Axion models such as \cite{ArkaniHamed:2003wu,ArkaniHamed:2003mz} attempt to suppress the symmetry breaking via the periodicity of the axion, which ensures periodic corrections as might arise from instantons. However, high scale inflation generated by such potentials requires a super-Planckian axion decay constant $f$.  Instanton corrections in known string theory examples become large \cite{Banks:2003sx}. Present indications are that such decay constants will not occur in a consistent theory of quantum gravity \cite{Banks:2003sx,ArkaniHamed:2006dz}.

The genericity of these dangerous operators is the root of the well-known ``$\eta$ problem" that plagues inflationary model building.  The difficulty in solving this problem in explicit string models has led to claims that the observation of tensor modes in the CMB would be inconsistent with string theory. However, several authors \cite{Silverstein:2008sg,McAllister:2008hb,Kaloper:2008fb,Berg:2009tg} have shown a possible way around the above difficulties. The basic mechanism is to take an axion with a {\it sub}-Planckian decay constant, and ``unwrap" it by coupling it to some additional degree of freedom. The unwrapping leads to a potential for the inflaton which dominates effects such as the periodic instanton-generated potentials expected in axion models.  The 4d field theory picture in \cite{Berg:2009tg}\ nicely illustrates how this can help: one spirals around a sub-Planckian patch of field space.  This is somewhat in the spirit of N-flation \cite{Dimopoulos:2005ac}, with a more explicit control for super-Planckian inflaton vevs. Nonetheless, as \cite{Silverstein:2008sg,McAllister:2008hb}\ found in the specific string models they examined, to achieve successful realization of these scenarios may require some work.

The major purpose of this paper is to study the viability of this class of inflation models from the standpoint of 4d effective field theory. We will do so via the 4d theory proposed in \cite{Kaloper:2008qs,Kaloper:2008fb}. The scalar acquires a mass via its mixing with a topological $4$-form, leading to an effective potential $V = \half \mu^2\phi^2$.  The Lagrangian in the axion-four-form sector has a global shift symmetry (we will explain why this does not violate Goldstone's theorem below), broken to a discrete shift symmetry by quantum effects when the gauge group is compact.  The energy as a function of the inflaton expectation value is a multibranched function so that while the energy of each branch is not periodic in $\phi$, the full theory is; this is related to the $\theta$-dependence of the energy of large-N QCD \cite{Witten:1980sp,Witten:1998uka}, which in fact gives a potential microscopic realization of our scenario.  We believe this model captures the physics of scenarios such as \cite{Silverstein:2008sg,McAllister:2008hb}, and that the model in \cite{Berg:2009tg}\ can be subsumed into this framework.

The inflaton will couple to other sectors as a pseudo-Nambu-Goldstone boson (pNGb) $\phi$.  Gauge and gravitational corrections can directly break the continuous shift symmetry, but these can be suppressed due to the periodicity of the axion. Additional corrections to the effective potential arise from corrections to the four-form Lagrangian.  Since the mixing between the axion and four-form is set by the inflaton mass $\mu = 10^{-6} m_{pl}$, the latter lead to corrections in the effective inflaton potential that are suppressed by powers of $(\mu/\Lambda)$ where $\Lambda$ is the fundamental scale of the theory, typically of the order of $10^{-2}\mpl$. 
One may think of the necessary smallness of $\mu$ as a fine tuning in the theory, although we expect that this small scale could arise from a dynamical mechanism. In fact, our model provides a hint of such a mechanism, since $\mu$ is really a landscape parameter, its value determined by the internal form fluxes of a compactification which may give rise to the low energy theory we employ. With the compactification dynamics controlled by GUT scale physics, achieving this scale seems plausible, and in the worst case, it is merely weakly anthropic.
Our real goal is to avoid having to fine tune the infinite number of couplings in the expansion $V = \sum g_n\phi^n/m_{pl}^{n-1}$, which would otherwise seem to be required for high-scale inflation to be viable.

In this work, we will focus on the original $\mu^2 \phi^2$ chaotic inflation scenario of Linde \cite{Linde:1983gd}. In principle, one can also realize the other linear or fractional power laws found in \cite{Silverstein:2008sg,McAllister:2008hb,Dong:2010in}.  We are also interested in the corrections,  which can be large enough to have interesting signatures in the CMB \cite{Flauger:2009ab, Hannestad:2009yx}, without spoiling slow-roll inflation.  After describing the general structure of corrections to the effective potential, we will describe the possible sources of such corrections arising from additional sectors that couple to the inflaton sector.  There are two especially stringent constraints.  The first is that moduli which couple to the inflaton must be comparatively heavy, with masses $M^2 \gg H^2$, where $H^2$ is the Hubble constant during inflation.  The second is that states which become light as the four-form becomes large do not push the gauge charge of the four-form up too high, although in the latter case high scale inflation models of the form constructed in \cite{Silverstein:2008sg,McAllister:2008hb}\ may still be viable.

In the next section we will review \cite{Kaloper:2008qs,Kaloper:2008fb} and provide a more detailed discussion of the underlying physics of the mass generation mechanism of those models.  In sec. 3 we will review quantum corrections from loops of the inflaton and graviton and remind the reader why these do not spoil slow-roll inflation.  In \S4 we will survey the possible corrections to these models that we expect to arise from integrating out additional degrees of freedom.  In section 5 we will discuss phenomenological constraints due to these corrections, and possible signatures.  In section 6 we will conclude.

\section{Chaotic inflation from $4$-form-scalar mixing}

The scenario proposed in \cite{Kaloper:2008fb} is defined by the following action:
\ba\label{eq:treeact}
S &=& \int d^4 x \sqrt{g}\, \left[ \frac{\mpl^2}{2} R - \frac{1}{2\cdot 4!} F_{\mu\nu\lambda\rho}F^{\mu\nu\lambda\rho} 
		- \half (\p\phi)^2 + \frac{\mu}{4!} \phi \epsilon^{\mu\nu\lambda\rho}F_{\mu\nu\lambda\rho}\right] ~~~~\nonumber \\
 & + & \frac16 \int d^4x \sqrt{g}\,\nabla_\mu \left[F^{\mu\nu\lambda\sigma} A_{\nu\lambda\sigma} - \mu \phi \frac{\epsilon^{\mu\nu\lambda\sigma}}{\sqrt{g}} A_{\nu\lambda\sigma} \right] \, .
\ea
The four-form $F = dA$ is the curl of a 3-form potential $A$, $F = dA$, and is invariant under the gauge transformation 
$A \to A + d\Omega$.  The boundary term on the second line will not be crucial for us, but is important in defining the theory \cite{Duff:1989ah, Duncan:1989ug}. It leads to a variational principle with fixed four-form field momentum at the boundary. The resulting path integral describes transitions between states with fixed field momentum  \cite{Duncan:1989ug}. Furthermore, in the presence of membranes appearing as charged $\delta$-function sources, with each side of the membrane taken as a spacetime-boundary, these boundary terms lead to the correct ``jump" of four-form flux across the membrane. The theory is gauge invariant up to boundary terms; for complete gauge invariance the gauge transformations must vanish at the boundaries or be canceled by the gauge transformation of charged sources at these boundaries. There is also an apparent global shift symmetry $\phi \to \phi + a$, whose fate we will discuss below.  

Eq. (\ref{eq:treeact}) is a direct 4d analog of 2d electrodynamics.\footnote{The relation between 2d electrodynamics and the 4d four-form theory was also pointed out in \cite{Seiberg:2010qd}.}  In two dimensions, the charged Dirac fermion can be bosonized, and the anomalous chiral symmetry of the fermions is dual to the shift symmetry of the boson  \cite{Witten:1978bc}.  In both two and four dimensions, the propagating degrees of freedom of this theory are known to be a massive scalar 
field \cite{Coleman:1976uz, Dvali:2005an, Kaloper:2008qs,Kaloper:2008fb}. To see this, let us compute the Hamiltonian.  
The fields with conjugate momenta are the scalar $\phi$ and the gauge field $A_{123}$:
\be
	q  \equiv p_{A_{123}} = F_{0123} - \mu\phi\ \, , ~~~~~~ \ \ \ \ \ p_{\phi} = \p_0 \phi \, .
	\label{mmenta}
\ee
We have denoted the four-form field momentum $q$ because it is identical to the Lagrange multiplier used in \cite{Kaloper:2008fb}. 
The Hamiltonian is:
\be\label{eq:treeham}
	H = \half (q + \mu\phi)^2 + \half p_{\phi}^2 + \half (\vec \nabla \phi)^2 \, ,
\ee
$q$ is conserved (the four-form has no dynamics by itself); $\phi$ behaves as a massive scalar with a minimum at $\phi = - q/\mu$.

Also as in two dimensions,\footnote{Again, see \cite{Seiberg:2010qd} for a useful review.} the physics and the unbroken symmetries of this model depends on the spectrum of charged sources, on the topology of the gauge group, and on the range of $\phi$.
Let us first consider the case that the 4-form gauge group is noncompact.  In this case, the Hamiltonian apparently has a symmetry under $\phi \to \phi + a$, $q \to q - \mu a$.  However, there is no massless Goldstone boson.  This is because the local current derived from Noether's theorem,
\be
	J = d\phi + *A\, ,
\ee
which is central to the current algebra proof of Goldstone's theorem,\footnote{See \cite{Weinberg:1996kr}\ for a review.}
is not invariant under gauge transformations of $A$, and so does not exist as a quantum operator. The
{\it integrated} charge $\int d^3 x J_0$, which generates global symmetries, does exist. Nonetheless, once $q$ is chosen the symmetry is broken by the dynamics. The scalar may or may not be periodic; since there is an operator $\int d^3 x A_{123}$ which shifts $q$, there is no obstacle to declaring that $\phi$ is periodic with any period desired\footnote{The massless mode of $\phi$ is removed by a gauge mode of $F$ in the equations of motion. We can see this by integrating $d^*d\phi = \mu F \equiv \mu dA$.  The right hand side is zero precisely when $F$ vanishes, so that $A$ is locally pure gauge, $A = d\Omega$.  Integrating the equation of motion, we find that $d\phi = \mu {}^*d\Omega$.}. Similarly, there is no restriction on the charge of any membranes. Depending on the membrane spectrum there can be multiple channels to reducing the energy for given values of $\phi$.    

When the four-form gauge group is compact, the shift symmetry is broken by quantum effects. The integrated current is not periodic and thus does not exist; $q$ is quantized and infinitesimal shifts do not exist. The membranes must have charges which are multiples of $e$, where $q$ is quantized in units of $e$. Now the scalar $\phi$ can be periodic, but the period $f_{\phi}$ of $\phi$ must satisfy the relation
\be\label{eq:axionrelation}
	\mu f_{\phi} = |e|\ ,
\ee
in order that a shift in $\phi$ can be compensated by a shift in $q$.  We will find that this is an important constraint.

Whether the gauge group and the range of $\phi$ are compact is, at this stage, a choice.  The choice should be guided by a UV completion.  There are strong arguments that in string theory, the four-form field strengths arise from compact gauge groups, and so are quantized \cite{Polchinski:1995sm,Bousso:2000xa}.  We will work within this scenario for the remainder of the present work.

The potential energy $\mu^2 \phi^2$ generated by the scalar is a result of spectral flow.  Consider the case of a compact gauge group with a nondynamical scalar appearing only as a ``$\theta$-term" $\mu \phi F^{(4)}$. If we start with $q = 0$, and increase $\phi$ at fixed $q$, the discrete energy levels begin to flow.  When $\phi$ is varied adiabatically to $\phi + |e|/\mu$, the ground state flows to become the first excited state; the excited state $q = - |e|$ flows to the ground state, and so on. Thus, the physical system is periodic in $\phi$ even if the energy of a given level is not, as demonstrated in Figure 1.  The result is a potential which can realize ``monodromy inflation" much as in \cite{Silverstein:2008sg,McAllister:2008hb}.

\begin{figure}[ht]
\centering
\epsfig{file=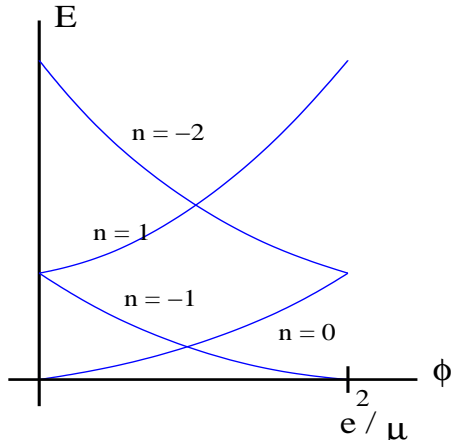,height=5.8cm,width=6cm}
\parbox{5.5in}{\caption{A map of the possible energies as a function of $\phi$, for the potential $V = \half(\mu \phi + q)^2$. The picture repeats itself (except for the labeling of the lines) each time one shifts $\phi \to \phi + |e|/m\equiv \phi + f_{\phi}$.}}
\end{figure}

\subsection{Monodromy from nonabelian gauge theory}

An axion coupled to the topological charge $\tr G \wedge G$ of pure $SU(N)$ Yang-Mills theory (where $G = dA + A^2$ is the nonabelian field strength) at large $N$ can also realize this phenomenon. As a function of $\theta$, Yang-Mills theory is thought to have a tower of vacuum states\cite{Witten:1980sp,Witten:1998uka}\  with energy \cite{Witten:1998uka}
\be\label{eq:scaxionpot}
	E_k(\theta) = N^2 h\left(\frac{\theta + 2\pi k}{N}\right)\, ;  \, \, \, \ k \in \Z \, .
\ee
At leading order in $1/N$ and at strong 't Hooft coupling $\lambda = g_{YM}^2 N$, Witten has shown that $h(x) = A x^2$, which leads to precisely the scenario discussed above. If we promote the theta angle to a propagating field, $\theta = \phi/f_{\phi}$, this is another realization of axion monodromy inflation.\footnote{One still needs to check if the theory in \cite{Witten:1998uka}\ will work for high-scale inflation. E.g., this is a dimensionally reduced 5d theory with a Kaluza-Klein scale of order the dynamical scale of the 4d gauge theory.}

As it has been pointed out in \cite{Di Vecchia:1980ve,Dvali:2005an}, the four-form theory can be used as the effective action for the coupled axion-gauge theory dynamics.  The Chern-Simons 3-form  $C \propto \tr \left( A d A - \frac{2}{3} A^3 \right)$ behaves as a massless 3-form field, with field strength $F = dC = \tr G \wedge G$. The physics above is reproduced by the Lagrangian ${\cal L} = \theta F + K(F)$ where $K$ is some unknown function.  The strong coupling result in \cite{Witten:1998uka}\ is consistent with a simple kinetic term $K(F) = F^2$, thus realizing our original model (\ref{eq:treeact}).

\subsection{Membrane nucleation and level crossing}

For monodromy inflation to work, transitions between branches of Fig. 1 should be suppressed.  Monodromy inflation will be safe if the lifetime for such transitions is long compared to the time scale of inflation. That is also necessary in order to avoid too large density perturbations, which could be $\sim {\cal O}(1)$ if the bubble nucleation rate is high. Since the nucleation processes are exponentially suppressed, we think that this is not too difficult to achieve. We will discuss some of the issues here, leaving a detailed analysis for future work.

Consider bubble nucleation in the flat space limit (valid if bubbles are smaller that $H^{-1}$ in size).  In the thin wall limit, the lifetime is proportional to \cite{Kobzarev:1974cp,Coleman:1977py,Callan:1977pt}:
\be\label{eq:decayrate}
	P = \lambda \exp \left(-\frac{27\pi^2}{2} \frac{\sigma^4}{(\Delta V)^3}\right) \, ,
\ee
where $\sigma$ is the domain wall tension, $\Delta V$ is the energy difference between the energy densities of the vacua separated by the bubble wall, and $\lambda$ is some scale arising from the fluctuation determinants around the classical instanton.  We will discuss the constraints and possible effects due to bubble nucleation in \S5.3.

\begin{figure}[ht]
\centering
\epsfig{file=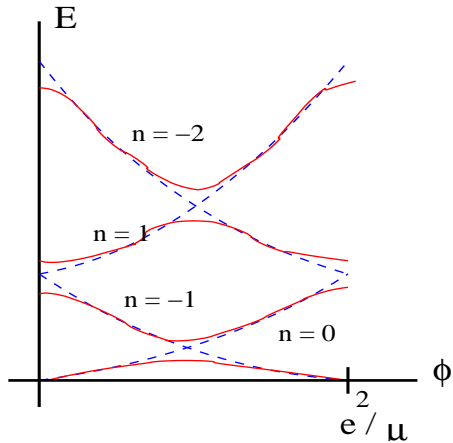,height=5.8cm,width=6cm}
\parbox{5.5in}{\caption{The solid red lines are the actual energies of the system as a function of the inflaton in flat space, when the Hamiltonian has matrix elements connecting adjacent values of $q$.  The dashed blue lines are the energy eigenstates when these matrix elements vanish.}}
\end{figure}

A related concern is that level repulsion due to some physics mixing the branches will split the level crossings at $\phi = (n + \half) f_{\phi}$ and the energy landscape as a function of $\phi$ will be deformed into finite, bounded bands, as seen in Fig. 2. If the inflaton evolved along these bands, it could spoil slow-roll inflation.  However, even in flat space at finite volume, the quantum state describing the time evolution of the inflaton will not follow the instantaneous energy levels if the transition time between different values of $q$ is large compared to the time scale of the evolution over a winding of the scalar. To get some intuition for this, we can consider a two-level system with Hamiltonian
\be
	H = \left( \begin{array}{ll} \hbar \alpha t & V \\ V^* &  - \hbar \alpha t \end{array} \right)\, .
\ee
It is easy to see that to leading order in $V$, the transition amplitude between adjacent values of $q$ is proportional to 
$|V|^2/\hbar^2\alpha^2$ as the system passes through the forbidden crossing region.  

The analysis of possible deformations is interesting, albeit involved. Here, we merely note that transitions via membrane nucleation occur through (sub) Hubble-sized bubbles \cite{Brown:1987dd,Brown:1988kg}.  Coherent fluctuations of the zero mode, beyond the first efolding that fluctuations are within our present horizon, will be forbidden by causality.  Within a Hubble volume, the scalar field could cease to go down along a single branch if the transition rate between adjacent values of flux was rapid over the time scale of a single winding of $\phi$.  This is safe if the bubble nucleation rate is small, which we must ensure at any rate.

\subsection{Tuning the mass, and quantum corrections}

In the next two sections we will address quantum corrections to (\ref{eq:treeact}). 
As we said in the introduction, one normally thinks of the selection of the mass parameter $\mu \sim 10^{-6}\mpl$ as a finely-tuned value.  In our setup, the specification of the numerical value of the mass may in fact be less severe than in the plain vanilla inflation models, since the mass $\mu$ is to be viewed as a landscape variable, a low energy consequence of some internal flux of a field that is integrated out after the compactification. However, this value can change from one location in the landscape to another, and scan the range of values around the required mass  $\mu \sim 10^{-6}\mpl$ that reproduced the observed amplitude of density fluctuations, if, as we have observed, the compactification dynamics is controlled by GUT scale physics. This means that this particular fine tuning may be relaxed somewhat, being determined by the weak anthropic principle.
The fine tuning we are concerned with is the more serious infinite fine tuning that simple chaotic inflation requires when the energy density is close to the upper bound -- any Planck-suppressed monomial in $\phi$ with ${\cal O}(1)$ coefficients spoils slow-roll inflation. A related statement is that the range of validity of 4d effective field theory is generically the Planck scale or lower.  Large-scale inflation takes one out of this range.

Monodromy provides a mechanism for the scalar field to travel over a large range without the 4d theory leaving its range of validity
\cite{Silverstein:2008sg,McAllister:2008hb,Kaloper:2008qs,Kaloper:2008fb, Berg:2009tg}.  The scalar $\phi$ makes many circuits over a region of size $f_{\phi} \equiv |e|/\mu \ll m_{pl}$.  The distance in field space between adjacent energy levels of the four-form sector may be even shorter than the periodicity of the scalar.  If the charged membranes are domain walls of a different scalar $\psi$ due to a potential $V(\psi)$ with multiple minima, the distance between adjacent local minima of 
$V(\psi)$ may be much shorter than $|e|/\mu$, realizing the pictures in \cite{Berg:2009tg}.

Corrections to the potential arise due to instanton effects, Planck-scale physics, and the coupling of other degrees of freedom such as moduli and Kaluza-Klein modes.  These corrections will be controlled by the compact topology of field space, and by the small parameter $\mu/\mpl$.  A caveat is that the required tuning of $\mu$ will require at least one additional fine tuning, due to (\ref{eq:axionrelation}).  The phenomenological requirement that $\mu \sim 10^{-6}$, combined with the conjecture that quantum gravity requires $f < \mpl$, means that $e \sim 10^{-3} \mpl$, which is already somewhat small. A large positive renormalization of $e$ could spoil this.  In this case it is possible that slow-roll inflation is viable, but that it is not controlled by the quadratic part of the potential, as in \cite{Silverstein:2008sg,McAllister:2008hb,Dong:2010in}. We will touch on the possibility of such models but not explore them too far in this work, since they require additional physical input (such as the validity of 10d supergravity) to be calculable.

\section{Perturbative quantum corrections from the inflationary sector}

\def\eps{{\epsilon}}

As we noted above, our principal aim in this paper is to understand the effects arising from coupling our chaotic inflation model to degrees of freedom which 
appear in a sensible UV completion.  To set the stage for this, let us start by reviewing the fact that within the inflaton-graviton sector,
perturbative corrections do {\it not} spoil slow roll chaotic inflation.  Readers already comfortable with this fact can skip to \S4.

The basic fact protecting the inflaton from quantum corrections is that slow roll demands that the inflaton potential be described by very small couplings. In other words, there is an approximate shift symmetry, as can be seen by studying the broken shift current:
\be
	\d \cdot J = V'(\phi) = \sqrt{\eps} \frac{V}{m_{pl}} = \sqrt{\eps} m_{pl} H^2 \ll m_{pl}^3\ , V^{3/4} \, ,
\ee
where $H^2 = V/m_{pl}^2$.  The symmetry breaking scale is smaller than the other physical scales in the problem ($V^{1/4}, m_{pl}$).

This approximate shift symmetry constrains quantum effects due to loops of inflatons and gravitons.
Since loop effects preserve the symmetry of the bare Lagrangian, symmetry-breaking loop effects must vanish when the symmetry-breaking terms in the bare Lagrangian vanish.  Furthermore, in models such as (\ref{eq:polynpot}), the breaking of the shift symmetry is small over a large range. Thus, we expect that the symmetry-breaking quantum corrections should not grow large with large $\phi$. In this section we will review the perturbative quantum corrections to chaotic inflation and stress 
these points in more detail.

UV dynamics -- nonperturbative effects, and couplings to other degrees of freedom -- are expected to break continuous global symmetries in theories that include quantum gravity. Nonperturbative gravitational effects can by themselves break shift symmetry \cite{Kallosh:1995hi}, even when the symmetry is present at tree level. This arises from loops of black holes and from Euclidean wormholes and instantons.  We will consider the latter two effects in the following section, together with the effects of integrating out other additional degrees of freedom.

\begin{figure}[ht]
\centering
\epsfig{file=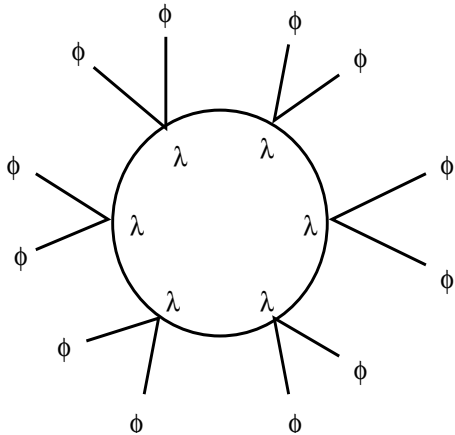,height=5.8cm,width=6cm}
\parbox{5.5in}{\caption{A "daisy diagram" that contributes to the one-loop effective potential.}}
\end{figure}

\subsection{Inflaton loops}

Let us consider $V \sim \lambda  \phi^4$ (although this is not actually the tree-level model we study in this paper). Slow roll requires that this potential remains flat for $\phi \gg m_{pl}$: $\partial^2_\phi V \ll m_{pl}^2, V/m_{pl}^2$. The one-loop correction is given by ``daisy" diagrams in quantum field theory, shown in Fig. 3.  While the potential contributes  $\delta m^2 \sim  \lambda \phi^2$, the curvature of the de Sitter background dominates the propagators, and we expect these diagrams to take the form
\be
V_{loops} = \sum_n c_n \lambda^n \frac{\phi^{2n}}{H^{2n-4}} = V \frac{V}{m_{pl}^4} \sum_n c_n \eta^n\, ,
\label{loops}
\ee
where $c_n$ are  ${\cal O}(1)$ numerical terms, and $\eta = m_{pl}^2 V''(\phi)/V\sim (m_{pl}/\phi)^2$ is the usual slow roll parameter. A more precise treatment of the one-loop correction in de Sitter space \cite{Dowker:1975xj} (with the mass in that paper set equal to $V''(\phi)$) indeed yields a potential of the form
\be
V_{loops} = H^4 F\left(\lambda \phi^2/H^2\right) = H^4 F(\eta) = V \left(\frac{V}{m_{pl}^4}\right)F(\eta) \, ,
\ee
where $F$ consists of positive powers and logarithms of its argument, and the term in parentheses is proportional
to the power in tensor modes, which is small.  Thus, we expect radiative corrections to the inflaton potential to 
be small out to extremely large $\phi$. Inflation is safe from inflaton loop corrections in the regime where it must match data.

\subsection{Graviton loops}

Generic operators in a UV-complete theory of quantum gravity are thought to spoil chaotic inflation \cite{Enqvist:1986vd}, absent some first-principles mechanism such as the one discussed in this paper.
However, Linde \cite{Linde:1987yb}, following the calculation in \cite{Smolin:1979ca}, pointed out that chaotic inflation is at least safe from perturbative quantum gravitational corrections in four dimensions. Let us sketch the one-loop calculation here.

The graviton-inflaton interactions are described by the terms:
\be
{\cal L}^1_{int} = \frac{1}{m_{pl}} \gamma_{\mu\nu} T^{\mu\nu} \, , ~~~~~~~~~~~~ 
{\cal L}^2_{int} = \frac{1}{m^2_{pl}} \gamma_{\mu\nu} \gamma^{\mu}{}_{\sigma} T^{\nu\sigma} \, , 
\label{vertex}
\ee
where $T_{\mu\nu}$ is the stress tensor for the inflaton:
\be
	T_{\mu\nu} = \d_{\mu}\phi\d_{\nu}\phi - \half g_{\mu\nu} \left[(\d\phi)^2 - V(\phi)\right] \, .
\ee
One-loop diagrams including gravitons in its internal legs will contribute terms which are powers of $V,V'$ and $V''$ suppressed by powers of the Planck mass; see \cite{Smolin:1979ca}\ for the full one-loop expression, including divergent contributions to the scalar potential. 

The finite terms are Planck-suppressed  {\it additive} corrections to the effective action, which come as the original potential or the curvature times positive powers of $V/m_{pl}^4$, $V'/m_{pl}^3$, and $V''/m_{pl}^2$. 
These will not spoil slow roll.  The divergent terms are:
\begin{itemize}
\item A quartic divergence which renormalizes the cosmological constant.
\item A quadratic divergences renormalizing $m_{pl}^2$.
\item A quadratic divergence proportional to $T/m_{pl}^2$.  
\item A logarithmic divergence proportional to $V''(\phi)$ renormalizing $m_{pl}^2$.
\item A logarithmic divergence proportional to $V^2/m_{pl}^4$ renormalizing the potential.
\end{itemize}
We will ignore the quartic divergence.  The quadratic divergence proportional to $T$ can be absorbed by a wavefunction renormalization of the metric.  The logarithmic divergence in the Planck mass is field-dependent.  If we perform a further field redefinition of the metric to keep the Einstein term canonically normalized, we get a logarithmically divergent term in the potential proportional to $V'' V/m_{pl}^2$.  However, when we cut the integral off at the Planck scale, the logarithimic correction is still only of ${\cal O}(1)$.
The result is \cite{Linde:1987yb}
\be
S_{eff} = \int d^4x \sqrt{g} \Bigl\{ \frac{m_{pl}^2 + m^2}{2} R - \frac12 (\partial \phi)^2 - V -
\Bigl(\frac{C_1 \partial^2_\phi V}{m_{pl}^2} + C_2 \frac{V}{m_{pl}^4} \Bigr) V \Bigr\} + \ldots \, , 
\label{renact}
\ee
where the dots denote the additional finite terms. If we start with a flat inflationary potential obeying $V \ll m^4_{pl}$ and $\mu^2 \ll m^2_{pl}$, the corrections are {\it small}. As one might have anticipated, the Smolin loops remain inconsequential, due to the perturbative shift symmetry.  The corrections are proportional to the scale of the shift symmetry breaking, and in the case of gravitational corrections are Planck-suppressed as well.

\section{Quantum corrections}
\label{quantumsection}

In this section we will consider the corrections to (\ref{eq:treeact})\ which we expect to arise from a UV-complete theory such as string theory. We will determine the hierarchies of scales required to maintain slow roll in the face of these corrections.  We will also consider the constraints imposed by moduli stabilization.  Our goal is to describe such corrections and constraints in as model-independent a fashion as possible, using the language of 4d effective field theory.   We will discuss detailed phenomenological constraints on the scales which appear, and possible observable effects, in \S5.

\subsection{Direct contributions to $V(\phi)$}

UV-complete quantum theories which include gravity are thought to break all continuous global symmetries.  Gravitational effects \cite{Abbott:1989jw,Barr:1992qq,Ghigna:1992iv,Kamionkowski:1992mf,Holman:1992us,Kallosh:1995hi} such as virtual black holes and wormholes \cite{Kallosh:1995hi}\ generically break such symmetries.  Anomalous couplings of the form $\phi \tr R \wedge R$ and $\phi \tr F \wedge F$ break the shift symmetry through gravitational and gauge instantons. In string theory, the symmetry may also be broken through string- or D-instantons. 

However, the fact that $\phi$ is a compact variable ensures that these corrections must be periodic with some period $f_{\phi}$.  Such potentials are typically bounded by some dynamical scale.  Furthermore, many axions in string theory are integrals of $p$-form gauge fields over $p$-cycles, so that the periodicity of the axion is the result of a discrete gauge invariance. Hence at distances shorter than the Kaluza-Klein scale, shifts of the axion become gauge modes \cite{Dimopoulos:2005ac}. This will cut off short-distance contributions to the direct potential for $\phi$.

Monodromy inflation gives a new wrinkle to axion-driven inflation models.  In previous models such as \cite{Freese:1990rb,ArkaniHamed:2003wu,ArkaniHamed:2003mz}, the inflaton potential itself is generated by such nonperturbative effects. One could try to keep them small to hold the inflaton in the slow-roll regime. However, single inflation models with super-Planckian field ranges would require a super-Planckian axion decay constant.  In known string constructions this leads to low instanton actions and thus the dominance of large instanton number, spoiling the scenario \cite{Banks:2003sx}. It has been argued that this is a general problem in any UV-complete theory of quantum gravity \cite{Vafa:2005ui,ArkaniHamed:2006dz}.

In the present case, as in \cite{Silverstein:2008sg,McAllister:2008hb,Kaloper:2008qs,Kaloper:2008fb}, we instead use the sub-Planckian periodicity of the axion to control the magnitude of direct corrections to the inflaton potential as the axion ranges over a super-Planckian distance. The reason that the {\it state} of the system is aperiodic as $\phi$ varies over a distance of $f_{\phi}$ is the spectral flow.

\subsubsection{Non-gravitational effects}\label{sss:inst}

Direct corrections to the potential of a periodic scalar typically come from instantons: they cannot be generated in perturbation theory if the underlying theory does not break the symmetry.  Such instantons can arise from non-Abelian gauge configurations, Euclidean string worldsheets, or Euclidean D-brane worldvolumes.  In cases that the dilute gas approximation is a good one, the potential is naturally written as:
\be\label{eq:periodiccorr}
	V_{corr}(\phi) = \Lambda_{uv}^4 \sum_n c_n \cos\left(\frac{n \phi}{f_{\phi}}\right) \, .
\ee
$c_n$ decreases with $n$ as $e^{-n S}$, where $S$ is the action of a single instanton.  Thus, the leading $n=1$ term dominates, and the height of the potential is approximately $\Lambda_{dyn}^4 = \Lambda_{uv}^4 e^{-S}$, which can be much smaller than the UV scale. The result is a trigonometric modulation of the $\mu^2\phi^2$ potential, as noted in \cite{McAllister:2008hb}. 

In string theory models the instanton action and axion decay constant are functions of moduli -- scalar fields which parametrize the compactification geometry, location of D-branes, and so on. We expect such additional degrees of freedom since interactions of a compact scalar field (to say nothing of gravity) are nonrenormalizable. A simple, renormalizable model which makes $f$ dynamical is the invisible axion of \cite{Dine:1981rt}.  Here the axion is the phase of a complex scalar field $\Phi$, whose modulus gets a vev from a potential such as $\lambda(|\Phi|^2 - v)^2$ which spontaneously breaks the shift symmetry. Cosine potentials arise from the terms:
\be\label{eq:insteft}
V_{corr} \sim \frac{g_{2n+m}}{\Lambda_{uv}^{2n+m-4}} |\Phi|^{2n} {\rm Re} \Bigl( \Phi^m \Bigr) \, .
\ee
In this case, $c_n \lesssim \left(f_{\phi}/\Lambda_{uv}\right)^n$. 

More generally, $f_{\phi} = {\rm const.}$ will locally define codimension-1 submanifolds in the moduli space, and we can use 
$\psi$ to coordinatize the normal direction to these submanifolds in a local patch of the moduli space (according to the moduli space metric).  Inverting the function $f(\psi)$, we can write the kinetic term as:
\be
	L = \half g(f) (\d f)^2 + f^2 d\theta^2
\ee
where $\phi = f\theta$.   These can be combined into a complex scalar, although this may or may not be a useful parametrization in effective field theory.

\subsubsection{Strong coupling effects}

When gauge theory instantons can become large and the theory is strongly coupled in the IR, the dilute gas approximation breaks down and the instanton expansion is poor.  Large-N analyses \cite{Witten:1980sp, Witten:1998uka}\ and lattice computations for pure $SU(3)$\cite{Giusti:2007tu}\ indicate that the potential is still periodic, with minima at $\phi = 0, \pm 2\pi f_{\phi}, \ldots$ and maxima at $\phi = \pm \pi f_{\phi}, \pm 3\pi f_{\phi}, \ldots$.  In these cases, the potential is much closer to that of (\ref{eq:scaxionpot}), with $h(x) \sim \Lambda_{dyn}^4 x^2$, where $\Lambda_{dyn}$ is some dynamical scale of the theory.\footnote{We thank R. Gopakumar and M. Marino for discussions on this point.} In terms of the effective field theory, corrections to the potential energy may be better approximated by powers of $\phi$ \cite{Witten:1998uka,Giusti:2007tu}
\be
	V \sim \Lambda_{dyn}^4 \sum_n \frac{\phi^n}{f_\phi^n}\, ,
\ee
than by powers of $f_\phi e^{i\phi/f_\phi}$ as in (\ref{eq:insteft}).

In these cases, the potential has multiple branches, just like the dominant four-form contribution to the potential.  
Simple chaotic inflation will work if the inflaton stays on the lowest-energy branch of the gauge-theory generated corrections.  At large $N$, the transitions between branches are highly suppressed, and these corrections become dangerous (although they could lead to a form of ``chain inflation" \cite{Freese:2004vs,Freese:2005kt,Freese:2006fk}).  We assume for the remainder of this paper that the direct corrections are of the type generated in \S\ref{sss:inst}.

\subsubsection{Gravitational corrections}

Nonperturbative gravitational effects -- gravitational instantons and virtual black holes -- break global symmetries.  
One class of gravitational instanton breaks the shift symmetry through anomalous couplings in the presence of topologically Euclidean solutions to the equations of motion.  With zero cosmological constant \cite{Holman:1992ah}, the instantons are compact Ricci-flat 4-manifolds such as K3 surfaces.  The topological term $\tr R {}^*R$ does not contribute to the anomalous conservation law for this shift symmetry, and the Einstein term vanishes. Gravitational instantons do contribute when there is a topologically nontrivial gauge field on this space.  The instanton action scales as $1/e_g^2$ where $e_g$ is the gauge field coupling, and one expects terms of the form $m_{pl}^4 e^{-8\pi/e_g^2}$.  This story must be modified during inflation, when the spacetime is approximately de Sitter and the instantons must satisfy Einstein's equations with a cosmological constant. But the instantons should generate terms of the form (\ref{eq:periodiccorr}).

The second class of instantons are wormholes which can break the shift symmetry directly \cite{Abbott:1989jw,Barr:1992qq,Ghigna:1992iv,Kamionkowski:1992mf,Holman:1992us,Kallosh:1995hi}. They will generate symmetry-breaking operators of the form (\ref{eq:insteft}), with UV scale $\Lambda_{uv}$ set equal to $\mpl$.   These effects were discussed explicitly in \cite{Kallosh:1995hi}, for axions with rigid decay constant and also for the case when the axion is the phase of a complex scalar.  The wormholes lead to exponentially suppressed $g_k$ when $\mpl r_0$ is large, where $r_0$ is the radius of the core of the wormhole throat.  Note that \cite{Kallosh:1995hi}\ discussed instantons in flat space.  Technically we should look for solutions in de Sitter space. Here we will focus only on the instantons which have throat sizes smaller than the Hubble scale, and so the corrections due to the background curvature may be small. 

In the simplest 4d models with the axion equal to the phase of a complex scalar $\Phi$, and a quartic symmetry-breaking potential for $\Phi$, the core of the wormhole reaches the 4d Planck scale \cite{Kallosh:1995hi}. In these cases the action is quite small, $S \sim - \ln (f_\phi/\mpl)$, and the symmetry breaking scale $f$ varies through the wormhole, reaching $f_\phi \sim m_{pl}$ in the throat.
Thus $e^{-S} \sim f_\phi/\mpl \sim 1$ in this case, and so the leading contribution is argued to be \cite{Kallosh:1995hi}
\be\label{eq:badop}
	\delta L \simeq \mpl^3 (\Phi + \Phi^*)\ , 
\ee
which is disastrously large. The discussion in \cite{Kallosh:1995hi}\ should be modified in the present case, due to the tree-level monodromy potential.ÊThe axion varies radially in these wormhole solutions, and one must take the potential into account in deducing the solution and the wormhole action, which we hope to address in more detail elsewhere. Ê

The "worst case" scenario (\ref{eq:badop}) seems to generate an axion potential which grows with $\mpl$. ÊThis seems odd if the main effect of large $\mpl$ is to decouple 4d gravity.ÊWe expect that the full story will actually be quite sensitive to the UV completion, especially since (\ref{eq:badop})\ arises from a particular class of potentials, from wormhole configurations with Planck-scale regions. Ê

For example the potential for the modulus controlling the axion decay constant could be quite rigid,Ê\cite{Kallosh:1995hi,Giddings:1987cg,Giddings:1988wv,Giddings:1989bq,Lee:1988ge}; in this case the wormhole throat is stabilized to a size well above $\mpl^{-1}$, the action is of order $\mpl/f_\phi$ and the symmetry-breaking effects are exponentially suppressed. ÊFurthermore, string-theoretic axions are often $p$-form gauge fields integrated over topologically non-trivial $p$-cycles in some compactification. ÊAt scales above the Kaluza-Klein scale, the gauge fields no longer sense the nontrivial topology and shifts at these distance scales
are gauge modes, which cuts off the effects of 4d gravity. ÊÊÊ

In this latter case, the weakest possibility is that all irrelevant operators are allowed. ÊThe leading term is
then:
\be
{\cal L}_{corr} \simeq {\cal C} \frac{f_\phi^5}{m_{pl}} \cos(\frac{\phi}{f_\phi}) \, ,
\label{logaction}
\ee
where ${\cal C}$ is a numerical factor. We will see in \S5\ that this will require a somewhat low axion decay
constant if $C \sim {\cal O}(1)$ or even $C \sim f_\phi/\mpl$.
However, we expect the suppressions to be even larger. Since the Kaluza-Klein scale $R_{KK}$ cuts off
gravitational contributions to the axion potential, the wormholes that do contribute may have throats with sizes
bounded at $R_{KK}$. ÊIn the examples in \cite{Kallosh:1995hi}\ in which the wormhole throat size was bounded
above $\mpl^{-1}$, the instanton Êaction typically had the form $e^{-c (\mpl r_0)^2}$, where $r_0$ is the size of
the wormhole throat, in the core of the wormhole. ÊEven a single order of magnitude difference between these
scales can lead to a large suppression.

\subsubsection{Maintaining slow roll}

Since we are working with $f_{\phi} < m_{pl}$, inflation should be driven by the potential due to $\phi$--$F$ coupling, and the instanton corrections should be subleading. Thus, $\Lambda \ll M_{gut}$, for corrections of the form  (\ref{eq:periodiccorr}).  Relatedly, for gravitational corrections of the form (\ref{logaction}), we require $f_\phi \ll M_{gut}^{2/3} m_{pl}^{1/3}$
which is easily realized for $f_\phi \le M_{gut}$.

A more stringent requirement on the instanton corrections is that the slow roll conditions $\epsilon \ll 1$ and $\eta \ll 1$ be preserved. This requires
\ba
\epsilon &=& \frac{3\dot \phi^2}{2V} = \frac{\mpl^2 V'^2}{2V^2} \ll 1 \, , \nonumber \\
\eta &=& \frac{\ddot \phi}{H \dot \phi} =  \frac{\mpl^2 V'^2}{V^2} - \frac{\mpl^2 V''}{V} \ll 1 \, .
\ea
We assume that $\Lambda^4 \ll \mu^2 \phi^2$, so that the dominant source of energy driving cosmic expansion comes from the $\phi$--$F$ mixing. We also require that and that $V'_{corr} < \mu^2 \phi$, so that the tree-level term controls the slow roll velocity $\dot \phi$, to ensue that the scalar density fluctuation is given by the standard formula
$\delta \rho/\rho \simeq H^2/\dot \phi$, with little variation over the course of inflation.  This will also ensure that the slope of the inflaton does not change sign during inflation.  Because $f_\phi < \mpl < \phi$, we average the trigonometric functions, to obtain
\be
\frac{\Lambda^4 \phi}{f_\phi} \ll  \mu^2 \phi^2 \, ,  ~~~~~~~~~~~~   \frac{\mpl^2\Lambda^4}{f_\phi^2} \ll \mu^2 \phi^2 \, .
\label{lambdabd1}
\ee
Substituting $\mu^2 \phi^2 \simeq M_{gut}^4$ which should be valid during the last ${\cal O}(60)$ efolds of inflation, 
we finally obtain
\be
\frac{\Lambda^4 \phi}{f_\phi} \ll  M_{gut}^4 \, ,  ~~~~~~~~~~~~   \frac{\mpl^2\Lambda^4}{f_\phi^2} \ll M_{gut}^4 \, .
\label{lambdabd2}
\ee
It is not hard to satisfy these constraints \cite{Kaloper:2008fb}. Depending on the ratio $\phi f_\phi/\mpl^2$, one is stronger than the other. For example, since $\phi$ needs to be ${\cal O}(10 \mpl)$ to produce ${\cal O}(60)$ efolds of inflation (see \S5), then $f_{\phi} \sim 0.1 m_{pl}$ implies $\phi f_\phi/ \mpl^2 \simeq 1$, and both inequalities give a similar bound $\Lambda \ll 0.3 M_{gut}$. On the other hand, if $f_{\phi} \sim M_{gut} \sim 10^{-2} \mpl$, then
$\phi f_\phi /\mpl^2 \simeq 0.1$, and the stronger bound comes from the second inequality (\ref{lambdabd2}), which yields
$\Lambda \ll 10^{15}\ GeV$. These scales are not difficult to realize in a natural way. For example, when the corrections arise from nonabelian gauge instantons,  $\Lambda$ is the scale at which a logarithmically running gauge coupling becomes of order $1$. If the gauge coupling starts even a little bit small at the fundamental scale $M_{gut}$, $\Lambda$ can be much smaller.

We note however that if the correction has larger derivatives than the tree level potential arising from $\phi$--$F$ mixing, then the smooth slow roll of the inflaton may be disrupted. Nevertheless inflation would still proceed as long as the total magnitude of the potential is small compared to the tree level term. In this case, the inflaton could even be classically trapped at each minimum, and inflation may realize a form of chain inflation \cite{Freese:2004vs}, somewhat analogously to \cite{Freese:2005kt}. 

\subsection{Indirect corrections}
 
Corrections to the dynamics of $F$, and derivative couplings of $F$ to $\d\phi$, lead to corrections to the effective potential during inflation, consistent with the periodicity of $\phi$. These corrections will be suppressed by powers of the small coupling $\mu$, and so are generically small once $\mu$ is generated well below the UV scale $\Lambda$.

First, consider corrections of the form
\be\label{eq:indirectcorr}
	\delta \cL = \sum_n c_n \frac{F^{2n}}{\Lambda^{4n - 4}}\, ,
\ee
where $\Lambda$ is some UV scale.   If we integrate out $F$, the effective potential takes the form
\be\label{eq:potexp}
	V(\phi) = \mu^2\phi^2 \sum_k d_k \left(\frac{\mu^2\phi^2}{\Lambda^4}\right)^k = V_0(\phi) \sum_k d_k
		\left(\frac{V_0(\phi)}{\Lambda^4}\right)^k \, .
\ee
The corrections to the slow roll parameters $\epsilon, \eta$ also take the form 
\be
	\delta (\eta, \epsilon) \sim \left( a \epsilon + b \eta\right) \left( \frac{V}{\Lambda^4} + \ldots \right) \, ,
\ee
where the dots denote higher orders in $V/\Lambda^4$.  Thus, the original potential $V_0(\phi)$ is a good approximation, and slow roll is maintained, if $\Lambda \gg V_0(\phi)^{1/4}$. 

Derivative couplings to $\phi$:
\be\label{eq:kincorr}
	\delta \cL \sim \frac{F^{2k}}{\Lambda^{4k}} (\d\phi)^2 + \ldots \, ,
\ee
lead to essentially the same constraints.  Here the dots refer to higher powers in both $F^2$ and $(\d\phi)^2$, and terms such as $(\d^2\phi)^2$. These corrections are small if $\Lambda \gg M_{gut}$.  They can also be recast as corrections to the potential. If we redefine $\phi \to \tphi$ where $\tphi$ is canonically normalized, the couplings to $(\d\phi)^2$ amount to a field-dependent renormalization of $f_{\phi}$; it will also lead to corrections to the effective potential of the form (\ref{eq:potexp}). 

\vskip .2cm
\noindent{\bf Summing up corrections to the potential}
\vskip .2cm

It is not necessarily a disaster for slow-roll inflation if the corrections (\ref{eq:indirectcorr}) dominate the leading potential.  In the original string theory models of monodromy inflation \cite{Silverstein:2008sg}, the inflaton potential takes the form $M_1^4 \sqrt{1 + \phi^2/M_2^2}$, $M_2 \ll m_{pl}$. Since $\phi \gg \mpl$ during inflation, the expansion in $\phi/M_2$ breaks down and the potential becomes linear. For the potential to behave approximately as $\mu^2\phi^2$, the scales must obey $M_2 \gg m_{pl}$, which is not realized in these string theory examples. High-scale inflation is still viable and the potential is calculable in these scenarios because the UV completion is understood: the large-$\phi$ behavior is controlled by 10d supergravity.  For large $\phi$, one can find potentials of the form $M^{4-\alpha} \phi^{\alpha}$ where $\alpha < 2$ can be a fraction \cite{Silverstein:2008sg,McAllister:2008hb,Dong:2010in}. 

A similar story holds for (\ref{eq:kincorr}), which has an equivalent effect upon redefining $\phi$ to be canonically normalized.   For example, consider the case that $d_k$ in (\ref{eq:potexp}) is small, and the $k=1$ term dominates (\ref{eq:kincorr}), with $\Lambda \leq M_{gut}$.  During inflation this ``correction" term controls the kinetic energy of $\phi$.  The canonically normalized scalar field is $\tilde \phi = \frac{\phi^2}{\Lambda}$, for which the potential is linear,  $V = \mu^2 \Lambda \tilde\phi$, along the lines of \cite{McAllister:2008hb, Takahashi:2010ky}. More generally, one needs control of the UV completion to sum the entire series in (\ref{eq:kincorr}).

\vskip .2cm
\noindent{\bf Large corrections to the kinetic term}
\vskip .2cm

When higher derivative terms $(\d\phi)^{2n} (\d^2\phi)^{n'} \ldots$ dominate\footnote{We view such operators as interactions, and disregard any new poles that such terms might appear to bring into the $\phi$-propagator -- among which may be ghosts -- by assuming these poles are above the cutoff.}, slow roll can be enforced by the kinetic terms, as in $k$-inflation \cite{ArmendarizPicon:1999rj}.  Non-canonical kinetic terms activated during inflation can provide for extra friction, as in the case of projectile ballistics at large velocities. Even if the corrections are small but not too small, there is still a chance for them to affect the scalar and tensor power spectrum  \cite{Kaloper:2002uj,Shiu:2002kg}. For example, if there is a term of the form $\frac{F^2}{\Lambda^6} (\d^2\phi)^2$, integrating out $F$ leads to a correction of size
\be
\delta \cL \sim \frac{F^2}{\Lambda^6} (\d^2\phi^2) \sim \frac{V(\phi)}{\Lambda^6} (\d^2\phi)^2 \sim \frac{1}{\Lambda_{eff}^2}
	(\d^2\phi)^2 \, .
\ee
If $\Lambda_{eff}^2 \gg H^2$, this will have a negligible effect on the CMBR \cite{Kaloper:2002uj}.  A sizeable effect would require 
$\Lambda \ll V^{1/4} = (F^2)^{1/4}$, in which case one must be able to sum the entire series in $F^2/\Lambda^4$.  This summation is possible in some string theory models; for example, when the inflaton corresponds to motion on the Coulomb branch of a strongly coupled gauge theory whose supergravity dual is known \cite{Silverstein:2003hf,Alishahiha:2004eh}.

\subsubsection{Examples of indirect corrections}

Corrections such as (\ref{eq:potexp}, \ref{eq:kincorr}) arise from integrating out degrees of freedom which couple to the inflationary sector.   The inflaton must couple to light degrees of freedom in order that it may decay and gracefully end inflation end inflation, reheating the universe in the process. These must have small couplings to the inflaton.  If we ignore the effects of the UV completion, the couplings can be protected by the same shift symmetry which protects inflation from inflaton and graviton loop corrections.

A UV-complete theory, however, will include couplings to massive fields.  Upon being integrated out, these can give dangerous corrections to the inflaton potential. In string theory, these fields include heavy moduli, Kaluza-Klein modes, and string modes.  Furthermore, there can be large numbers of such modes below the UV cutoff. There are often hundreds of moduli in a string compactification, and order $V_D m_*^D = (\mpl^2/m_*^2)$ Kaluza-Klein modes of a given field in a D-dimensional compactification with volume D and UV cutoff $m_*$.  We would like to estimate the effects of integrating out these modes, in terms of the effective field theory for $\phi, F$.  The guiding principle is that the discrete shift symmetry forbids couplings of powers of $\phi$ 
to additional fields; typical couplings to heavy states will be to powers of $F$ and to derivatives of $\phi$.

\vskip .2cm
\noindent{\bf Integrating out moduli}
\vskip .2cm

Explicit string theory models tend to have scalars which are light compared to the Kaluza-Klein scale (see \cite{Kachru:2006em}\ for a fairly general discussion). These scalars parameterize the geometry of the underlying compactification manifold, and the positions of D-branes.  Closed string moduli are further distinguished in that they have $\mpl$-suppressed couplings to other fields in the 4d effective theory.  The fact that they are generally light makes them especially dangerous for high-scale inflation, when they couple to the inflationary sector. (In \S4.3 we will discuss the requirement that inflation not destabilize the moduli.)

Consider a ``radion" modulus  $\psi$, related to the physical size of an extra dimension by
\be
	L(\psi) = L_0 e^{\psi/\mpl} \, ,
	\label{radion}
\ee
The canonically normalized field $\psi$ is stabilized at $\psi = 0$ by a mass term $\half M^2\psi^2$. Direct couplings of $\psi$ to $\phi$ must respect the periodicity of $\phi$. We can also have couplings to powers of $F^2$ and $(\d\phi)^2$.  Such couplings can lead to inflation-driven shifts of the moduli vevs and moduli masses.  The former will lead to tree-level corrections to the inflaton dynamics; the latter will lead to one-loop corrections following the calculation of  \cite{Coleman:1973jx}. Note that the shifting moduli vevs will also shift the masses of Kaluza-Klein modes; we will discuss this below.

If $\psi$ is canonically normalized, its couplings to 4d fields are suppressed by the 4d Planck scale $\mpl$. Thus, the leading order coupling is  
\be
	\delta \cL = - \frac{\psi}{\mpl} \left(a (\d\phi)^2 + b F^2 + c \Lambda^4 \cos (\phi/f_{\phi}) \right)\ldots \, .
	\label{radcoupling}
\ee
During inflation, both $(\d_t\phi)^2$ and $F^2$ have classical expectation values; these induce a tadpole for $\psi$ via (\ref{radcoupling}):
\be\label{eq:modshift}
	\delta\psi = \frac{1}{\mpl M^2} \left(a (\d\phi)^2 + b F^2 + c \Lambda^4 \cos\left(\frac{\phi}{f_{\phi}}\right)\right) \, .
\ee
The third term is already constrained to be subleading by the results of \S4.1\ and we will ignore it from here on out.
Plugging (\ref{eq:modshift}) back into (\ref{radcoupling}) and into the mass term for $\psi$, we find:
\be
	\cL_{eff} = - \frac{1}{2 \mpl^2 M^2} \left( a^2(\d\phi)^2 + 2ab (\d\phi)^2F^2 + b^2 F^4 \right) \, .
\ee
In deducing corrections, note that $\vev{F^2} \sim V$ and $\vev{(\d_t\phi)^2} \sim \epsilon V$, where $\epsilon$ is the slow-roll parameter.  The upshot is that we get corrections of the form (\ref{eq:potexp},\ref{eq:kincorr}) with $\Lambda^2 \sim m_{pl} M$.
In this case $V \ll \Lambda^4 \Rightarrow H^2 \ll M^2$, where $H^2$ is the Hubble scale during inflation.  Thus, the moduli give small corrections to $V_{tree}$ only if they are comparatively heavy. In \S4.3\ we will find that moduli stabilization will require the same inequality be satisfied. In this case, if many moduli couple to the inflaton and four-form, these effects will be enhanced and the constraint on the moduli masses will be stricter.

The shift in the modulus also leads to a shift in the compactification radius: 
\be
	\frac{\delta L}{L} \sim \frac{\delta \psi}{\mpl} \sim \frac{H^2}{M^2} \, ,
\label{eq:decomp}
\ee
which is again small when $H^2 \ll M^2$. This will correct the 4d Planck scale:
\be
	\delta m_{pl}^2 = m_*^{4+D} \delta V_D \sim m_{pl}^2 \frac{H^2}{M^2} \, ,
\ee
where $D$ is the dimension of the compactification manifold, and $m_*\gtrsim M_{gut}$ the higher-dimensional Planck scale.  Note that the slow roll parameters $\epsilon, \eta$ have quadratic dependence on the Planck scale; the correction above will lead to a multiplicative shift of the slow roll parameters of order $H^2/M^2$, which imposes the same constraint on $M$ as above. 

The heuristic picture is that for $M < H$, $\psi$ will move during inflation due to classical forcing and quantum fluctuations in the nearly de Sitter background.  This adds kinetic energy to the universe and prevents the inflaton from driving vacuum-dominated cosmic expansion. The motion could even start to decompactify hidden dimensions.  Making the moduli heavier than the Hubble parameter during inflation, $M > H$, prevents this.

Quadratic couplings to the modulus will also generate corrections.  Consider a coupling 
\be 
	\delta \cL \sim - c \frac{(\psi - \psi_0)^2}{\mpl^2} F^2 \, ,
\ee
where $c \sim \cO(1)$. At tree level, integrating out $\psi$ leads to corrections proportional to $\psi_0/\mpl$, $H/M$.  The latter constraint has already appeared; the former constraint requires that the modulus not be pushed Planckian distances. 
See \cite{Dong:2010in}\ for further discussions of this kind of coupling\footnote{Note, however, that we are restricting our attention to the ``universally" coupled moduli with 4d Planck suppressed couplings, instead of the more general, stronger couplings considered in \cite{Dong:2010in}. If the couplings are stronger than gravitational by a factor of $g$, one finds stronger bounds, $M > g H$.}.
Since the moduli mass is shifted to $M_{\psi}^2 = M^2 + c \frac{V}{\mpl^2}$, integrating out $\psi$ at one loop leads to a Coleman-Weinberg potential for $\phi$ of the form 
\begin{eqnarray}
	M_{\psi}^4 \ln M_{\psi}^2 & = & \left(M^2 + c \frac{V^4}{\mpl^2}\right)^2 \ln \left(M^2 + c \frac{V^4}{\mpl^2}\right)\\
	& = & \left(M^2 + c H^2\right)^2 \ln\left(M^2 +  c H^2\right) \, .
\end{eqnarray}
If $M^2 < H^2$, the leading correction is $\cO(H^4) \sim V (V/\mpl^4) \ll V$ and is harmless; other constraints rule out this mass range for moduli. If $M^2 > H^2$, the leading non-constant term proportional to $M^4$ is $M^4 (V/\mpl^4)$.  This leads to a multiplicative renormalization of $V$ of order $M^4/\mpl^4$. The terms proportional to $M^2$ take the form
\be
	\frac{\delta V}{V} \sim \frac{2 c M^2 + c^2 H^2}{\mpl^2} \ln\left(M^2 + V/m_p^2\right) \, .
	\label{potsshift}
\ee
Even hundreds of moduli, all of these corrections are small, especially when $M \lesssim M_{gut}$.

Note, that when integrating out the modulus {\it does} substantially change the inflaton potential, often the potential becomes flattened, maintaining the slow-roll even better \cite{Dong:2010in}. This has been noted in the explicit string theory examples in \cite{Silverstein:2008sg, McAllister:2008hb}.

\vskip .2cm
\noindent{\bf Integrating out Kaluza-Klein modes}
\vskip .2cm

Integrating out Kaluza-Klein modes $\chi_n$ for gravitons will also yield shifts of the effective potential.  There are two effects leading to $F$-dependent masses for $\chi$.  First, there can be direct couplings of the form:
\be
	\delta \cL = - d \frac{\chi^2}{\mpl^2} F^2 \, .
\ee
Secondly, the shift (\ref{eq:decomp}) changes the compactification radius.

Since the Kaluza-Klein modes are generally heavier than the moduli, we will integrate them out first.
For a $D$-dimensional compactification, the 4d effective potential at one loop, arising from integrating out all of the Kaluza-Klein modes, will be:
\be\label{eq:oneloopkk}
	\cL_{eff} \sim a V_D(\psi/\mpl) \int d^4k  d^D q \ln \left(k^2 + q^2 + d \frac{F^2}{m_p^2} + \ldots \right) \, .
\ee
Here $a$ is a prefactor which includes the sum over graviton polarizations and phase space factors. The latter typically provide inverse powers of $2\pi$ and such which can render $a \sim \cO(1)$ (see for example \cite{Kaloper:2002uj}\ for a related discussion).   $k$, $q$ are the momenta along $\R^4$ and the compactification directions, respectively. $\cV_D$ is the compactification volume, which depends on the moduli $\psi$.  We are schematically representing the sum over modes as $V_D \int d^D q$, which is somewhat justified since the dominant effects will come from the high momenta near the cutoff.  

We take this cutoff to be the $D+4$-dimensional Planck scale $m_*$.  In principle we should compute the integral (\ref{eq:oneloopkk})\ via an honest heat kernel expansion, but we can identify how the different terms depend on the scales of the problem.  The leading divergence will be:
\be\label{eq:oneloopmp}
	V^{(1)}(\psi/\mpl) = a m_*^{D+4} \cV_D(\psi) \, ,
\ee
and will give a $\phi$-independent contribution to the moduli potential. We will absorb this into the moduli potential.

The heat kernel expansion will give further corrections as a power series in the effective mass $F^2/\mpl^2$ and the $D+4$-dimensional curvature, with the corresponding lower powers of the cutoff dictated by dimensional analysis.  The next most divergent pair of terms after (\ref{eq:oneloopmp}) are quadratic in $F$ or linear in the $D+4$-dimensional Ricci scalar. For the $F^2$ term, near the minimum of $\psi$ we find
\be\label{eq:kineticren}
	\delta \cL \sim a  m_*^{D+2} \cV_D(\psi) \frac{F^2}{\mpl^2} 
	\sim a \left(1 + \cV_D'(\psi) \frac{\delta \psi}{\mpl} + \ldots \right) F^2 \, .
\ee
The first term can be absorbed into the normalization of $F$ and renormalizes $e$. Since $\mu f_{\phi} = |e|$ and $\mu$ is fixed by observational considerations, one should take care that this does not force $f > \mpl$; so long as $a \sim \cO(1)$ this is not an immediate concern. The second term in (\ref{eq:kineticren}) is a term of the form (\ref{radcoupling}).  Higher-order terms in $\delta\phi$ can be written using (\ref{eq:modshift}) as a power series in $F^2/(\mpl^2 M^2)$, so that we get corrections of the form (\ref{eq:indirectcorr}) with $\Lambda^2 = \mpl M$.  These lead to multiplicative renormalizations of the tree-level potential with size $H^2/M^2$. Once again, we find that we must impose the constraint $H \ll M$

The term linear in the Ricci scalar $\cR_{D+4}$ includes the 4d Einstein term. Inserting the shift in the moduli in it gives a multiplicative renormalization of Newton's constant by terms of the form $\sum_{k=0} c_k \frac{H^{2k}}{M^{2k}}$ with $c_k \sim \O(a)$, which is again safe if $M^2 \gg H^2$. The term linear in the curvature of the compactification manifold will renormalize the moduli potential and kinetic energy by similar terms.

The additional terms are of higher order in $F^2/(m_*^2 m_{pl}^2) \sim H^2/m_*^2$ and in $\cR/m_*^2$.  We expect them to lead to small corrections when $H^2 \ll M^2, m_*^2$ and $V < m_*^2$.

\vskip .2cm
\noindent{\bf Additional light states}
\vskip .2cm

\begin{figure}[ht]
\centering
\epsfig{file=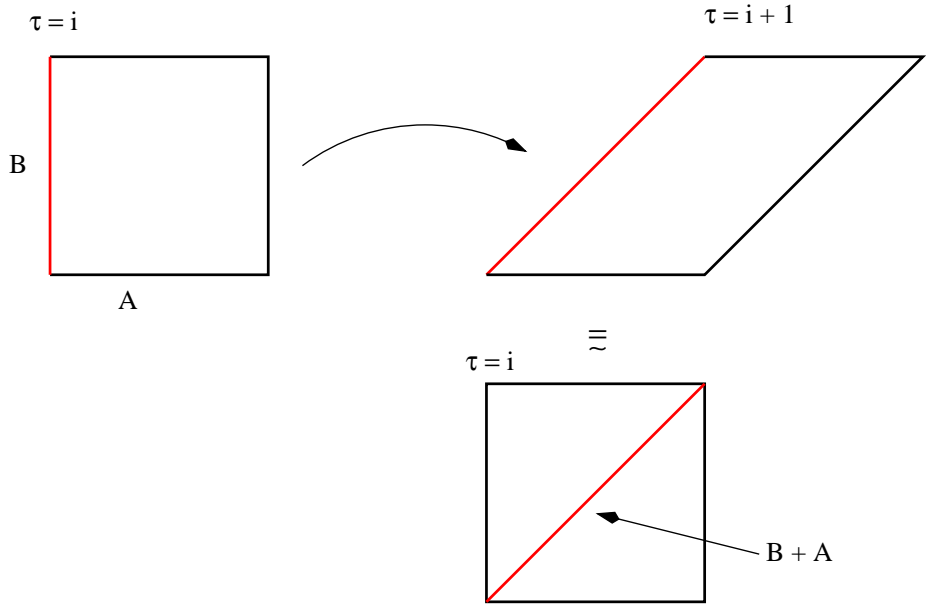,height=8cm}
\parbox{5.5in}{\caption{The effects of a shift of the complex structure $\tau \to \tau + 1$ of a 2-torus, in the presence of a D-brane wrapping a cycle of the torus.}}
\end{figure}

\begin{figure}[ht]
\centering
\epsfig{file=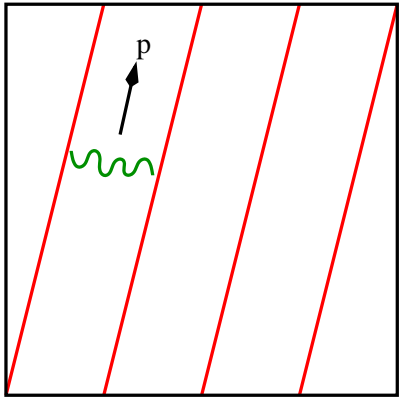,height=5.3cm}
\parbox{5.5in}{\caption{Light states in the presence of a multiply wrapped D-brane. If the torus is a square with sides of length $L$, and the brane wraps the cycle $A + n B$, the short stretched string shown will have length $L/\sqrt{1+n^2}$.  The total length of the wrapped brane is $L \sqrt{1 + n^2}$, so the lowest momentum mode will have mass $1/(L\sqrt{1 + n^2})$.}}
\label{tiltedtorus}
\end{figure}

In string-theoretic models of axion monodromy inflation, new light states appear as the axion winds many times  \cite{Silverstein:2008sg, McAllister:2008hb}.   As an example, consider the complex structure modulus $\tau$ of a $T^2$ factor of the compactification. $\tau$ is a periodic variable under a shift $\tau \to \tau + 1$.  Begin with $\tau = i$ on a square torus with length $L_0^2$, and wrap a D4-brane along the imaginary direction, as shown in in Fig. (\ref{tiltedtorus}).  Shifts of the complex structure $\tau \to \tau + n$ will cause the D4-brane to acquire winding along the cycle $A$: the D-brane will wrap the cycle $B + n A$ and will increase in energy. 
As the D4-brne winds, light string states appear: open strings stretching between leaves of the winding, and momentum modes of open strings which effectively live on a 1d space with length $L_0\sqrt{1 + n^2}$.  Simple embeddings of this model  do not give viable models of inflation \cite{Silverstein:2008sg, McAllister:2008hb}.  However, the model is illustrative of the sorts of effects we must worry about, so we will use it as a benchmark.

After $n$ windings, the open string theory is essentially a $U(n)$ gauge theory.  The effects on inflation depend qualitatively on whether the 't Hooft coupling $g_s n$ is weak or strong.  In the weakly-coupled theory, the spectrum of momentum and winding modes is identical to that of Kaluza-Klein mode on an asymmetric torus with sides $L_1 = L \sqrt{1 + n^2}$, $L_2 = (\alpha'/L) \sqrt{1 + n^2}$, and volume $\ell_s^2 (1 + n^2)$.  The loop will be cut off at most at the string scale if not below.\footnote{Technically one must take care to cancel the closed-string tadpoles.} It is easy to see that there are order $n^2$ modes below the cutoff, consistent with a weakly-coupled $U(n)$ gauge theory.

Let us consider an effective field theory of our inflaton-four form theory coupled to a set of light states that contains the basic features listed above. First, the coupling of these light states to the inflaton sector should maintain the periodicity of $\phi$.  We can do this by coupling the open string modes to $F^2$ in such a way that they become light as one increases $F$. Let us assume that the light modes are labeled by  quantum numbers $n_{k = 1, \ldots, p}$ (these could be momentum or winding) and mass scales $M_k$ (scales associated to momentum or winding). Then, the candidate (canonically normalized) Lagrangian for light scalar modes is:
\be
	\cL_{\psi_{n}} \sim \half (\d\psi_n)^2 - \sum_k n_k^2 M_k^2 \left(1 - a \frac{F^2}{\Lambda^4} + \ldots\right) \psi_n^2 \, ,
\ee
where the higher order terms are higher powers in $F^2$. The effective number of such modes below the cutoff $\Lambda_{light}$ will be:
\be
	N_{eff} \sim \tV_p \left( 1 + \sum_{n=1} c_n \frac{F^{2n}}{\Lambda^{4n}}\right) \, ,
\ee
where $\tV_p \equiv \frac{\Lambda_{light}^p}{M_1 \ldots M_p}$ is an effective volume in units of the cutoff. In the example above, $\Lambda^2 = e$ (so that $F^2/\Lambda^4 = n^2$), $\Lambda_{light} = m_s$, $m_1 = 1/(L\sqrt{1+n^2})$, and $m_2 = L/(\alpha'\sqrt{1+n^2})$.

Now we can integrate out the light modes.  We will focus on two leading effects.  First, the Planck scale will be renormalized.  We will use the heuristic formula:
\be
	\delta \mpl^2 = N_{eff} \Lambda_{light}^2 \, .
\ee
Then
\be\label{eq:weakplanckcorr}
	\frac{\delta \mpl^2}{\mpl^2} \sim  \frac{\Lambda_{light}^2}{\mpl^2}\tV_p \left( 1 + \sum_{r=1} c_n \frac{F^{2r}}{\Lambda^{4r}}\right) \, .
\ee
The basic requirement for slow roll to work is that this shift is small during inflation.  The first prefactor is small, of order $\lesssim 10^{-4}$.  $\tV_p$ depends on the model, but in the example above it is of order 1.  The sum is more problematic.  In the example above, $F^2/e^4 \sim n^2$, where $n$ is the ``winding number" of the axion.  For example when $f \sim 0.1 \mpl$, $n = 100$ and the $r = 1$ term is already $10^{4}$ by design.  This can lead to an order 1 multiplicative renormalization of the Planck scale.  If $\tV_p$ or $c_1$ is large, this already becomes problematic; furthermore, the series expansion will break down badly.
Thus, in the weak coupling case, we expect we will need $\Lambda^2 \gg e = \mu f_{\phi}$.  As an example, $\mu \sim 10^{-6} \mpl$, and $f \sim 0.1 \mpl$, implies $e \sim (3\times 10^{-4} \mpl)^2 < M_{gut}^2$, so this does not seem unattainable in principle.  We also remind the reader that this example is known to fail at any rate, and that the weak 't Hooft coupling assumption $g_s n \ll 1$, with $g_s$ the string coupling, means that $g_s \ll 10^{-2}$.  

Secondly, there will be a correction to the vacuum energy arising from these light modes.  The details of this correction depend on the spin of the modes as well as the coupling.  At one loop, a crude estimate of the leading term is
\be\label{eq:vacenergy}
	E \sim \Lambda_{light}^4 \left(N_{eff,0}(F^2/\Lambda^4) - 2 N_{eff,1/2}(F^2/\Lambda^4) + \ldots\right) + \ldots \, ,
\ee
where $N_{j,eff}$ is some effective number of real light modes with spin $j$. The sign and magnitude will depend on details of the spectrum, the nature of the cutoff, and so on. When $N_{eff,j}$ is large we can imagine performing a large-N expansion; the leading terms will be $\cO(N_{eff})$.  In the example above, $N_{j,eff} \sim F^2/\Lambda^4 \sim n^2$, and (\ref{eq:vacenergy}) corresponds to a wavefunction renormalization for the four-form.  If $\Lambda_{light} > \Lambda$, this is a substantial correction.  The corrected potential becomes:
\be
	V = \frac{(\mu \phi - k e)^2}{1 - a\frac{\Lambda_{light}^4}{\Lambda^4}} \, ,
\ee
where $a$ is some model-dependent constant.  If the denominator becomes sufficiently negative, the theory runs the risk of becoming destabilized.

At strong 't Hooft coupling, we should calculate the effects using a supergravity dual.  In a full string theory, this amounts to correctly accounting for the backreaction of the four-form on the compactification. Much of this is taken care of in our earlier discussion of the effects of moduli and Kaluza-Klein states.  We will focus on the possibility that large four-form flux induces a localized region of strong warping, in which many degrees of freedom reside, as happens if large D-brane charge is induced.

Let us imagine that flux $F\sim n e$, built up by $n$ windings of the inflaton, leads to a strongly-coupled 4d $U(n)$ gauge theory localized somewhere in the compactification.  The gauge dynamics can be described by a Randall-Sundrum-type "throat region" \cite{McAllister:2008hb,Randall:1999ee,Randall:1999vf}.  We will assume the supergravity parameters are related to the gauge theory parameters just as they are in the $N=4$ supersymmetric theory; this is not realistic, but gives us a handle on the size of the effects involved.

The curvature radius of the throat is $R \sim \lambda^{1/4} \sqrt{\alpha'}$, 
where $\lambda$ is the 't Hooft coupling.  The correction to the Planck scale is $\delta \mpl^2 \sim m_{pl,5}^3 R$ \cite{Randall:1999ee,Randall:1999vf}, where $m_{pl,5}$ is the 5d Planck scale down the throat.  Using $n^2 = (m_{pl,5} R)^3$, we find that $\delta \mpl^2 = \frac{n^2}{R^2}$, consistent with a field theory with $n^2$ degrees of freedom and a cutoff $1/R$.  This is a reasonable value of the cutoff; it is the scale at which the throat opens up into the compactification geometry, and at higher energies excitations leave the throat and live in the bulk of the $D$-dimensional compactification. The fractional shift in the 4d Planck scale is:
\be
	\frac{\delta \mpl^2}{\mpl^2} \sim \frac{n^2}{\lambda^{1/2}} \frac{m_s^2}{\mpl^2} \, .
\ee
If $n\sim 100$, $\lambda \sim 10$, and $m_s < M_{gut} \sim 10^{-2} \mpl$, this is $< 0.3$ and is not a significant shift.  (Recall that $\delta \eta \sim \eta (\delta m_{pl}^2/m_{pl}^2)$.) However, if $n$ increases by a factor of 2, $\delta \mpl^2/\mpl^2 \sim \cO(1)$.  Thus these parameters are on the border of respectability.  As a related check, the throat is a small perturbation of the geometry if the Randall-Sundrum throat size is smaller than the Kaluza-Klein scale.  This requires $\lambda^{1/4} \ell_s < \ell_{KK}$.  If $\lambda \sim 100$, this merely requires $\ell_{KK} \sim 3 \ell_s$.

We should similarly compute renormalizations to the $F$-dependent part of the Lagrangian. The 4d vacuum energy contrbuted by the throat will be of order 
\begin{eqnarray}
	E_{light} & \sim & \frac{m_{5,pl}^3}{R} \sim \frac{n^2}{R^4} \sim \frac{n^2}{\lambda} m_s^4\\
	&  \sim & \frac{n m_s^4}{g_s} \sim n m_*^4 \, ,
\end{eqnarray}
where $m_* = g_s^{1/4} m_s$ is the 10d Planck scale. If $*F = n e$, then this energy is linear in $F$, and dominates the tree-level inflaton action for $V_{tree} \ll m_*$.  A more realistic variant of this scenario could lead to some of the $\phi^{\alpha < 2}$ inflaton potentials studied by \cite{Silverstein:2008sg, McAllister:2008hb,Dong:2010in}.  To preserve quadratic inflation, we must avoid such a dramatic enhancement of the number of degrees of freedom, or ensure that their effects are strongly cut off.

\subsection{Moduli stabilization}

Another constraint \cite{Silverstein:2008sg,McAllister:2008hb} arises from the fact that the coefficients $c_n$ of  higher-dimension operator corrections to the tree-level $\phi-F$ potential generically depend on moduli in string theory.  Consider a particular modulus $\psi$.  The combined potential is:
\be
	V(\phi,\psi) = V_{mod}(\psi) + \half \mu\left(\psi\right)^2 \phi^2 + c_1\left(\psi\right) \Lambda^4 \cos\left(\frac{\phi}{f_{\phi}}\right) \, .
\ee
The cosine term changes in sign many times during inflation; the leading inflaton potential also sources a contribution to the moduli potential during inflation.  Thus, $\psi$ will be destabilized unless $V_{mod}'' \gg |c_1''| \Lambda^4/\mpl^2, |(\mu^2)''|\phi^2/\mpl^2$. 
This is a serious (though not necessarily insurmountable) constraint in existing compact string models with moduli stabilization \cite{Silverstein:2008sg,McAllister:2008hb}. The constraint can be satisfied if $\psi$ is stabilized classically or perturbatively; or if it is stabilized by an instanton coming from a sector which does not couple to $\phi$ and whose dynamical scale is much larger than $\Lambda$. 

To get a feel for these constraints, let $V_0 \sim \half M^2 \psi^2$ and $c_1 = c_1(\psi/m_{pl}), \mu^2 = \mu^2(\psi/\mpl)$, as would be expected if $\psi$ is a closed string modulus.  If $c_1$ is a dimensionless function with magnitude and derivatives of order ${\cal O}(1)$, the instanton corrections will not destabilize the moduli if $M^2 \gg \Lambda^4/m_{pl}^2$. Similarly, if $\mu^2(x) = \mu_0^2 k(x)$ with $k(0) = 1$ and derivatives of $k$ of ${\cal O}(1)$, then the the moduli are not destabilized so long as $M^2 \gg \mu_0^2 \frac{\phi^2}{m_{pl}^2} \sim H^2$.  This is the same constraint we found by demanding that integrating out the moduli does not induce too-large corrections to inflation.

\subsection{Dual formulation of corrections}

We close this section by noting that we can also write the UV corrections such as (\ref{eq:potexp})\ in terms of a 2-form gauge field dual to the axion, following \cite{Dvali:2005an}.  Dual $p$-form models and their effective theories have also been discussed previously in \cite{fern}, in connection with the effective field theory description of defect condensates.

Consider the following action (ignoring the coupling to gravity),
\be
S = - \int d^4 x \sqrt{g} \Bigl\{\frac{1}{2\cdot 4!} F_{\mu\nu\lambda\sigma}^2 + \frac{1}{2 \cdot 3!} 
\Bigl(\mu A_{\mu\nu\lambda} - H_{\mu\nu\lambda}\Bigr)^2 \Bigr\} \, ,
\label{gaugeforms}
\ee
where $H_{\mu\nu\lambda} = 3! \partial_{[\mu} B_{\nu\lambda]}$. Here $B, A$ are dimension-1 gauge fields and $\mu$ is a dimension-1 gauge coupling.  This theory is manifestly invariant under the gauge symmetry
\be
A_{\mu\nu\lambda} \rightarrow A_{\mu\nu\lambda} + \partial_{[\mu} \Omega_{\nu\lambda]} \, ,
~~~~~~~~~ B_{\nu\lambda} \rightarrow B_{\nu\lambda} + \mu \Omega_{\nu\lambda} \, ,
\label{formgaugetrafo}
\ee
where $\Omega_{\nu\lambda}$ are the components of an arbitrary $2$-form. This gauge symmetry will be preserved by gravitational effects.

As pointed out in \cite{Dvali:2005an}, the theory (\ref{gaugeforms}) has a single massive scalar degree of freedom. We can see this by adding the term
\be\label{eq:bianchilm}
	\delta L_{Bianchi} \sim \frac{\zeta^2}{24} a *\left(dA - F \right)\ ,
\ee
which enforces $F = dA$. Here  $a$ is a dimensionless, periodic scalar with period $2\pi$, and $\zeta$ a normalization factor.  The periodicity enforces quantization of $F$.  Upon integrating out $A$, we find that $\frac{\zeta^2}{\mu} da = *(H - eA)$. Inserting this into (\ref{gaugeforms},\ref{eq:bianchilm}), we find the action (\ref{eq:treeact})\ with $\phi = \frac{\zeta^2}{\mu} a$, and $\zeta = e$.  Thus the mass term in (\ref{eq:treeact}) is dual to a gauge coupling, and the periodicity of $\phi$ is related directly to the flux quantization of $F$.

Higher-dimension corrections are constrained by gauge invariance and locality to take the form:
\be
S = - \int d^4 x \sqrt{g} \Bigl\{ \Lambda^4 P\Bigl(\frac{F_{\mu\nu\lambda\sigma}^2}{\Lambda^4}\Bigr) + \Lambda^4 Q\Bigl[\frac{\Bigl(\mu A_{\mu\nu\lambda} - H_{\mu\nu\lambda}\Bigr)^2}{\Lambda^4} \Bigr] \Bigr\} \, ,
\label{gaugeformscorr}
\ee
where $\Lambda$ is an effective UV cutoff, and the $\mu$-dependence is fixed by the demand that the weak gauge coupling limit $\mu \to 0$ is smooth. $P(x)$ and $Q(x)$ are analytic functions starting with a linear terms $Q = x + \ldots$, $P = x + \ldots$. Naturalness states that the coefficients of the Taylor expansions for $P$ and $Q$ are of order unity\footnote{Note, that the normalization of these terms is an {\it assumption}; the tree level terms can be renormalized, such that $Q = Z_Q x + \ldots$, $P = Z_P x + \ldots$. The renormalization factors $Z_Q, Z_P$ will in general depend on the UV realizations of the theory, and if their ratio is large the mass $\mu$ can get renormalized too, as we have been exploring in detail so far. If there are mechanisms to fix this ratio to be of order unity, a small $\mu$ is technically natural.}, after absorbing wave function renormalization factors away. Writing out the first few terms of this series, we find that:
\beqa
S &=& - \int d^4 x \sqrt{g} \Bigl\{\frac{1}{2\cdot 4!} {F}_{\mu\nu\lambda\sigma}^2 
+ \frac{c_1}{\Lambda^4}  {F}_{\mu\nu\lambda\sigma}^4 +  \frac{c_2}{\Lambda^8}  {F}_{\mu\nu\lambda\sigma}^6  + \ldots  \nonumber \\ 
&& ~~~~~~~~~~~~ + \frac{1}{2 \cdot 3!} 
\Bigl(\mu { A}_{\mu\nu\lambda} - H_{\mu\nu\lambda}\Bigr)^2 + \frac{d_1}{\Lambda^4} 
\Bigl(\mu { A}_{\mu\nu\lambda} - H_{\mu\nu\lambda}\Bigr)^4 + \ldots \Bigr\} \, ,
\label{gaugeformscorrcan}
\eeqa
where $c_i$ and $d_i$ are all $O(1)$ numbers.

Upon adding the term (\ref{eq:bianchilm})\ and integrating out $A$, we find that higher-order terms in $Q$ lead to terms of the form
$(\d\phi)^{2 + 2k}/\Lambda^{4k}$, while $P$ corresponds to the terms in (\ref{eq:indirectcorr}).  Thus we can recast our entire discussion in terms of a 2-form gauge field dual to the scalar axion.  Furthermore, we can describe multiple effects by replacing $A$ with a sum of 3-forms $A_k$, and $\Lambda^4 P(x)$ by $\sum_k \Lambda_k^4 P_k(F_k^2\Lambda_k^4)$.  This can capture multiple physical effects, including the ``direct corrections" in \S4.1 \cite{Dvali:2005an}. 

This formalism may help find a mechanism for generating naturally small masses for scalar degrees of freedom.  While the masses appear to be protected by global continuous symmetries in perturbation theory, quantum gravity seems to forbid global symmetries. If the symmetries can be reinterpreted as local symmetries gauged by $p$-form gauge fields, there is a possible route towards explaining small scalar masses \cite{Dvali:2005an}.

\section{Phenomenological features and constraints}

In this section we give a quantitative discussion of the quantum effects which are generic to our field theory model.  We will open with constraints which arise if we demand that the corrections do not spoil slow roll inflation controlled by the leading $\phi$--$F$ potential.  In the following discussion we will review the phenomenon of "resonant non-Gaussianity" which appears due to sinusoidal modulations of the potential\cite{Chen:2008wn,Flauger:2009ab,Flauger:2010ja}. In the final section we will discuss modes of nonperturbative instability in our models, and describe a possible signature due to jumps in $\mu$.

We will focus on corrections to slow roll inflation driven by a quadratic potential.  We will put aside for now the flatter potentials described in \cite{Silverstein:2008sg,McAllister:2008hb,Dong:2010in}.

\subsection{Constraints from the slow roll criterion}

The scales which appear in our model are: the mass $\mu$, the axion decay constant $f_{\phi}$, the various UV scales governing the $F^n$ and higher-derivative corrections to the axion kinetic term, the masses $M$ of moduli coupling to the inflaton, and the dynamical scale of gauge groups which couple to the inflaton via the topological term. Here we will summarize the numerical bounds on these scales which must be obeyed to yield viable inflation. 

First, the mass of the inflaton $\mu$ is fixed by the observed power in CMB fluctuations. The density perturbations produced by the inflaton are $\frac{\delta \rho}{\rho} = \frac{\mu}{4\sqrt{6} \pi m_{pl}} (\frac{\phi}{m_{pl}} )^2 = \frac{\mu}{\sqrt{6} \pi m_{pl}} {\cal N}$, where ${\cal N} = \frac14 (\frac{\phi}{m_{pl}} )^2$ is the number of efolds before the end of inflation where the density contrast is calculated. COBE gives the following constraint on fluctuations which entered the horizon, that are usually normalized to the standard pivot at $50$ efoldings before the end of inflation:
\be
\frac{\delta \rho}{\rho} \simeq 6.5 \times \frac{\mu}{ m_{pl}} \sim 5\times 10^{-5}\, . \label{contrastnum}
\ee
The inflaton mass is therefore:
\be
\mu \simeq 0.77 \times 10^{-5} m_{pl} \simeq 1.5 \times 10^{13} \, GeV \, .
\label{massnumerics}
\ee

Second, let us consider the bounds on the various UV scales $\Lambda$. In all of the examples given in \S4, the corrections to the inflaton potential were terms of the form $V (V/\Lambda^4)^k$; so we demand that $V \ll \Lambda^4$. Given $\mu$, to estimate $V$ we must compute the magnitude of $\phi$ during inflation.  For inflation to produce at least $\sim 65$ efolds of inflation, this value of mass must extend for field values of at least 
\be
\phi_* \sim 2 \sqrt{\cal N} m_{pl} \simeq 14 \mpl \simeq 3.4 \times 10^{19} \, GeV \, ,
\label{vevvalue}
\ee
The corresponding energy scale is:
\be
V^{1/4} =  \sqrt{\frac{1}{\sqrt{2}}\mu \phi_*} \simeq 1.9 \times 10^{16} \, GeV \, ,
\label{cutoff}
\ee
which is on the order of the GUT scale, $V^{1/4} \sim M_{gut} \sim 2 \times 10^{16}\ GeV$, at which the gauge couplings seem to unify in the MSSM \cite{Dimopoulos:1981yj}. Hence the flatness of the $\mu^2 \phi^2$ potential in our scenario can be safe from quantum corrections due to physics above the GUT scale. In many string-theoretic GUT models, the fundamental scale is the 10d Planck scale which is somewhat above the GUT scale.  So meeting the bound $\Lambda > M_{gut}$ does not seem out of reach.

The moduli constraint from section (\ref{quantumsection}) implied that the moduli mass $M$ needs to be at least bigger than the Hubble scale during inflation,
\be
H  \simeq \frac{\mu \phi}{\sqrt{6} m_{pl}} \simeq 0.9 \times 10^{14} \,  GeV\, .
\label{eq:hubblescale}
\ee
This is a nontrivial constraint.  For example, for moduli stabilized by $p$-form magnetic fluxes in 10d string compactifications with Kaluza-Klein scale $R = M_{KK}^{-1}$, the moduli masses are typically $M \sim \alpha'/R^3 = M_{KK} (\alpha'/R^2)$ \cite{Kachru:2006em}.  If $M_{KK} \sim M_{gut}$ (required for 4d unification), this means that the string scale should be within an order of magnitude of the Kaluza-Klein scale. Moduli stabilized by non-peturbative effects such as string or D-brane instantons are more problematic; their masses die off exponentially as powers of $R^2/\alpha'$ \cite{Kachru:2006em}. Such moduli run the risk of becoming completely destabilized if they couple to the inflaton sector. 

As we discussed in \S4.1.1, instanton effects generate corrections to the potential of the form (\ref{eq:periodiccorr}). These are expected in generic string compactifications. Furthermore, couplings of the inflaton $\phi$ to nonabelian gauge groups of the form $\frac{\phi}{f_\phi} Tr(G ^*G)$ ensure the end of inflation and subsequent reheating, while not breaking shift symmetry perturbatively. In order to maintain slow-roll inflation driven by the tree-level quadratic potential, we must place bounds on the dynamical scale $\Lambda_{dyn}$, setting the height of the instanton-generated potential,\footnote{Again, we are restricting ourself to the case that the instanton expansion is good.} and the axion decay constant $f_{\phi}$. For the energy in the gauge instantons to be subdominant, $\Lambda_{dyn}$ must be below $V^{1/4}$. In general, $\Lambda_{dyn}$ is related to a UV scale $\Lambda$, such as the  string or 10d Planck scale, by an exponential of the instanton action $e^{-S}$. In the case of gauge instantons, the scale is generally the scale at which the gauge coupling of an asymptotically free gauge theory becomes strong, after being run down from this fundamental scale. If $\Lambda$ also appears as in Eqs. (\ref{eq:indirectcorr},\ref{eq:kincorr}), then there are competing constraints as $\Lambda$ must be well above the scale of the inflation potential.  Thus the instanton action must be large, and the gauge couplings at scale $\Lambda$ small, in order to generate a substantial hierarchy
\be\label{eq:scalehier}
	\Lambda_{dyn}^4 \sim \Lambda^4 e^{-S} \ll V \ll \Lambda^4 \, .
\ee
For example, one may get sufficient numerical range of parameters with $S\sim 18$, which will generate $\Lambda_{dyn}/\Lambda \sim 10^{-2}$, and then fit the scale of inflation in between.

The requirement that the slow roll parameters remain small puts a bound on $f_\phi$, via (\ref{lambdabd1},\ref{lambdabd2}); for the cases of interest to us, this translates to roughly
\be\label{eq:adchier}
	m_{pl} > f_\phi \gg \mpl \frac{\Lambda_{dyn}^2}{V^{1/2}} \, .
\ee
The first inequality arises from the conjecture that super-Planckian decay constants do not exist in quantum gravity \cite{Banks:2003sx,ArkaniHamed:2003wu}. The second constraint means that $f_{\phi}$ must be somewhat high.  Using (\ref{vevvalue}), we note first that for the two inequalities to be compatible, $\Lambda_{dyn} < \half V^{1/4}$, consistent with (\ref{eq:scalehier}).  As a guide, if we set $f_\phi = 0.1 \mpl$, and use (\ref{vevvalue}), this requires $\Lambda < 0.3 M_{gut}$.  For fixed $\Lambda_{dyn}$, (\ref{eq:adchier}) can be viewed as a bound on the number of windings of the axion during inflation, $N_w \equiv \phi_*/f$:
\be
	N_w \ll \sqrt{\frac{2}{\eps}} \frac{V^{1/2}}{\Lambda_{dyn}^2} \, .
\ee
In our model, $\eps = \half \mpl^2 \left(\frac{V'}{V}\right)^2 = 2 \frac{\mpl^2}{\phi_*^2} \sim 10^{-2}$.  Thus, if there is an order of magnitude separation between $V^{1/4}$ and $\Lambda_{dyn}$, $N_w \ll 1400$. 

We have less understanding of nonperturbative gravitational effects, but we will survey some possibilities.  When the inflaton is the phase of a complex scalar, the worst case scenario \cite{Kallosh:1995hi}\ is that wormholes generate terms of the form (\ref{eq:periodiccorr}) with $c_1 \Lambda_{uv}^4 \sim m_{pl}^3 f_\phi$, corresponding to the correction (\ref{eq:badop}). Using (\ref{eq:scalehier}), this requires $f_\phi \ll V/\mpl^3$.  On the other hand (\ref{lambdabd1},\ref{lambdabd2}) requires $f_\phi^2 \gg \frac{\mpl^2 \Lambda^4}{V} \sim \mpl^2 \frac{m_{pl}^3 f_\phi}{V}$.  The compatibility of these two inequalities would require $V \gg \mpl^4$, so this correction must not occur.

If we take the point of view that symmetry breaking effects should vanish in the limit $\mpl \to \infty$ (this is supported by the idea that the axion emerges from some higher dimensional $p$-form with a gauge symmetry), then if the instanton action is still a logarithm of $f_\phi/\mpl$, the next worst case would be that wormholes generate irrelevant operators
\be\label{eq:okirrel}
	V_{grav} = \frac{1}{\mpl^{p - 3}}|\Phi|^{p\geq 4}(\Phi + \Phi^*) \, ,
\ee
with coefficients of order unity. For $p = 4$, this is equivalent to (\ref{eq:periodiccorr}) with $\Lambda_{dyn}^4 = c_1 \Lambda_{uv}^4 = \frac{f_\phi^5}{\mpl}$. Then eq. (\ref{eq:adchier}) yields $f^2_\phi \gg \mpl^2 \frac{f_\phi^5}{\mpl V}$, or therefore
$f_\phi \ll (V/\mpl)^{1/3} \sim 3.8 \times 10^{15} \ GeV$. This is a little low, but not impossible. If on the other hand the leading operator has $p = 5$, then (\ref{eq:adchier}) reduces to $f^2_\phi \gg \mpl^2 \frac{f^6_\phi}{ \mpl^2 V}$, which yields 
$f_\phi < V^{1/4} \sim 1.9 \times 10^{16} \ GeV$, quite a plausible scenario.

However, it is likely that the wormhole effects are much weaker, as we discussed in \S4.1.3.  In fact, already a rigid potential for the axion decay constant leads to a suppression for these effects of order $g\sim e^{- \mpl/f_\phi}$; for $f \sim 0.1 \mpl$, this is of order $4.5\times 10^{-5}$. For the worst case scenario (\ref{eq:badop}) this does not help, but for the operators (\ref{eq:okirrel}) our model is in very good shape.  Furthermore, if in addition the Kaluza-Klein scale cuts off the wormholes, as one may argue when the axion comes from some $p$-form gauge theory, the instanton action could be as high as $\mpl^2 R_{KK}^2$.  If $R_{KK} \geq M_{gut}$, then $S > 10^4$ and these effects are completely invisible, and the nonperturbative gravitational effects are totally harmless.

Before we close this section, let us also note that inflation driven by a quadratic potential has the best shot at generating detectable relic gravitational waves. Quartic potentials are all but ruled out by present data, and potentials with lower powers of the inflaton as in \cite{Silverstein:2008sg,McAllister:2008hb,Dong:2010in}\ produce a smaller tensor-to-scalar ratio.  Tensor fluctuations are proportional to the Hubble scale during inflation in Planck units, $\delta \rho/\rho|_T \sim H/m_{pl}$, and for inflation driven by a quadratic potential the Hubble scale is as large as can be. More precisely, using the Lyth consistency equation \cite{Lyth:1996im,Efstathiou:2005tq} relating the tensor density contrast to $\phi/\mpl$ during inflation, we find: 
\be
r \simeq 10^{-2} \Bigl(\frac{\phi}{\sqrt{8\pi}m_{pl}}\Bigr)^2 \simeq 0.08 \, .
\label{rratio}
\ee
This is within reach of future cosmological probes of relic gravitational waves.

\subsection{Periodic corrections to the potential}

As we discussed in \S4.1, our inflationary model can receive direct periodic corrections to the tree-level axion potential, through nonperturbative effects.  There and here, we consider the cases where these corrections can be successfully computed by an instanton expansion in the dilute gas approximation.  The leading order direct correction is then a small oscillating modulation of the inflaton potential:
\begin{equation}
\delta V(\phi)=\Lambda^4\,\cos\left(\frac{\phi}{f_\phi}\right) \, .
\label{cospots}
\end{equation}

The correction (\ref{cospots})\ generates oscillatory features in the power spectrum.  If $\dot\phi\gg H\,f_\phi$, the period is small compared to a Hubble time. For sub-horizon modes of the right frequencies, the two- and three-point functions of the inflaton can become resonant, leading to an enhancement of the power spectrum and the bispectrum \cite{Chen: 2008wn,Flauger:2009ab,Flauger:2010ja}.  These effects, and the window for sub-Hubble oscillations of $V$, are enhanced at smaller $f_\phi$. Note, however, that smaller $f_\phi$ requires  larger number of windings over inflation, $N_w \sim 10 m_{pl}/f_\phi$.  Sufficiently large $N_w$ can lead to problematic corrections from light states, as described at the end of \S4.2.

The standard gauge $f_{NL}$ of non-Gaussianity is schematically 
\be\label{eq:fnlschem}
	f_{NL}=\langle \zeta^3 \rangle/\langle \zeta^2\rangle^2\, ,
\ee
where $\zeta$ is the gauge invariant density perturbation. The precise computation in  \cite{Maldacena:2002vr,Chen:2006xjb,Chen:2008wn,Flauger:2009ab,Flauger:2010ja}\  is somewhat complicated. As we are interested in order-of-magnitude effects, we now give a crude sketch of their derivation which highlights the essential physics of this phenomenon, and gets the answer in \cite{Flauger:2010ja}\ up to a numerical prefactor independent of the parameters.

When $\phi = \phi(t)$, the correction (\ref{cospots})\ leads to a potential which oscillates in time.  The basic phenomenon of resonance occurs when the combined frequencies of $\zeta$ in the two- or three-pont functions are in sync with this resonance.  The quantum wavefuncitons of the inflaton are oscillatory in time only when the physical momenta are below the Hubble scale.
Thus, resonance will occur when the period $\dot\phi/f$ of the oscillations (\ref{cospots})\ is less than $H^{-1}$. 

Because the resonance affects modes only while they are sub-horizon, we can replace the curvature perturbation $\zeta$ by the inflaton fluctuation $\delta\phi$, to which its dynamics reduces at short distances, with the relationship $\zeta\sim H\,\delta\phi/\dot\phi$.  To leading order, when the spectrum is nearly scale invariant, the two-point function is $\langle \zeta\zeta\rangle \sim H^4/\dot\phi^2$.  Note that the two-point function itself will get corrections due to resonance; we will argue that these are small.
The gauge $f_{NL}$ is thus 
\be 
	f_{NL} \sim \frac{\dot{\phi}}{H^5}\langle\delta\phi(x_1)\,\delta\phi(x_2)\,\delta\phi(x_3)\rangle\ .
\ee
which follows from (\ref{eq:fnlschem})\ if we use the tree-level potential $V = \half \mu^2\phi^2$ to derive $\langle \zeta^2\rangle$.

The inflaton 3-point function can then be computed via perturbative quantum field theory, with the perturbations proportional to the slow roll parameters and their derivatives.  The full interaction Hamiltonian must be derived by properly fixing the diffeomorphism invariance and expanding in small fluctuations in the metric and the scalar.  In general this requires care, as shown in \cite{Maldacena:2002vr}.  For example, in the gauge in which the scalar fluctuations are parametrized by the inflaton fluctuation $\delta\phi$, one might expect the leading term cubic in fluctuations is proportional to $V'''\delta\phi^3$.  For single-field slow-roll without sinusoidal perturbations, this is not the case.  However, one can see from \cite{Chen:2006xjb,Chen:2008wn,Flauger:2009ab,Flauger:2010ja}\ that this term does contribute at leading order to the resonant non-Gaussianity. 
The leading contribution to the three-point function is thus:
\begin{equation}
\langle\delta\phi^3\rangle\simeq \int \prod_i d^3{\bf k}_i\,e^{i\sum \bf{k}_i\bf{x}_i}\delta^{(3)}\left(\sum {\bf k}_i\right)\prod_i\delta\phi_{{\bf {k}}_i}(t)\int \frac{dt'}{(H t')^4} V'''(\phi(t'))\prod_i\delta\phi^*_{-{\bf {k}}_i}(t')+{\mathrm {h.c.}} \,,
\label{3pt}
\end{equation}
where the time $t$ is conformal time, and the momenta $k$ are the comoving momenta. (Note that $\dot\phi$ denotes the derivative with respect to proper time.) After a mode has crossed the horizon, its amplitude is $\delta\phi_{\bf k}\simeq H\,e^{-ikt}/k^{3/2} \simeq H/k^{3/2}$, the horizon crossing time for this mode being defined as $k t \sim 1$. We use this approximate expression for the factors of $\delta\phi$ outside of the time integral (the ones arising from the operators insertions).  Before horizon crossing, the quantum modes behave as $\delta\phi_{\bf k}\simeq (H t) e^{-ikt}/k^{1/2}$. The time-dependent prefactor simply signifies that amplitude of subhorizon modes redshifts with the cosmic expansion, behaving as the inverse of the wavelength. This can be inferred by using the virial theorem for subhorizon harmonic oscillators in de Sitter space. Since resonance occurs when the modes are inside the horizon, we use this expression for the factors of $\delta\phi$ inside the integral, which come from the interaction Hamiltonian.  Thus, the integrals in (\ref{3pt}) reduce to 
\begin{equation}\label{eq:ftpint}
\langle\delta\phi^3\rangle\simeq H^3\,\int \left\{\prod_i\frac{d^3{\bf k}_i}{k_i^2}\,e^{i\sum ({\bf{k}}_i\,{\bf{x}}_i-k_i\,t)}\delta^{(3)}\left(\sum {\bf k}_i\right)\right\}\int \frac{dt'}{Ht} V'''(\phi(t'))e^{i\sum k_i\,t'}+{\mathrm {h.c.}}\,.
\end{equation}
The factor of $t$ in the time integral is {\it not} integrated over. It just encodes the scale at which the sum of frequencies $\sum k_i$ comes into resonance with the sinusoidal potential.

The $\delta$-function in (\ref{eq:ftpint})\ imposing momentum conservation means that the vectors $\vec{k_i}$ form a triangle; the expression in curly brackets determines the shape of the dominant triangles.  Denote the coefficient of the term in curly brackets as 
$\overline{\langle\delta\phi^3\rangle}$.  For a given shape, the numerator in (\ref{eq:fnlschem}) is defined as 
\be
	\langle \zeta^3\rangle = \left(\frac{H}{\dot\phi}\right)^3 \overline{\langle\delta\phi^3\rangle} \, .
\ee
Finally, inserting the leading expression $\langle \zeta^2\rangle \sim H^4/\dot{\phi}^2$ into the denominator of (\ref{eq:fnlschem}), we find:
\begin{equation}
f_{NL}\sim \frac{\dot\phi}{H^3}\int \frac{dt'}{t'}\, V'''(\phi(t'))e^{i\sum k_i\,t'}\, .
\label{fnl}
\end{equation}

This integral can be evaluated by the saddle point method.  Resonance occurs over a short physical time scale, so we can approximate $\phi(t) = \phi_0 + \dot{\phi} t_{phys} = \phi_0 + \frac{\dot\phi}{H} \ln (- t H)$, where $t_{phys}$ is the physical time.  $\phi_0$ gives an overall phase which we ignore, so that:
\begin{equation}\label{fnlshort}
V'''(\phi(t))\sim\frac{\Lambda^4}{f_\phi^3}\cos\left(\frac{\dot\phi}{Hf_\phi}\ln (- H t) \right) \, .
\end{equation}
Separating the cosine into two exponentials, the dominant contribution to (\ref{fnl}) comes from a Gaussian integral centered on
$t_{res} = - \dot\phi/(Hf_\phi |k_{res}|)$, with width $\Delta t = \sqrt{f_\phi |k_{res}| H/\dot{\phi}}$, where we use the shorthand $k_{res} = (\sum k_i)_{res}$. Doing this integral, we find:
\begin{equation}
f_{NL, res}\sim \Lambda^4 \sqrt{\frac{\dot\phi}{H^5\,f_\phi^5}}\,\,.
\label{nongauss}
\end{equation}
up to numerical factors.  This has the same parametric dependence as the complete result in \cite{Flauger:2010ja}. For times close to $t_{res}$, the proper time is $t_p = t_{conf}/(H t_{res})$ and the physical momentum is $k_{phys} = k H t_{res}$.  Thus, we can write the "resonance condition" as $k_{res,phys} = \dot\phi/f_\phi$.

The condition for the physical time scale over which resonance occurs to be short compared to a Hubble time is:
\be
	H \Delta t_{phys} = \frac{\Delta t}{t_{res}} = \sqrt{\frac{H f_\phi}{\dot\phi}} \sim \sqrt{\frac{f_\phi}{\mpl \sqrt{\eps}}} < 1 \, .
\ee
For the quadratic model with $\eps \sim 10^{-2}$,  this is achieved for $f_\phi \simeq 0.1 \mpl$.

The instanton potential~(\ref{cospots}) also leads to an oscillating correction to scalar power spectrum ${\cal P}$~\cite{Flauger:2010ja}. The amplitude of the correction is
\begin{equation}
\frac{\Delta{\cal P}}{{\cal {P}}}\simeq
\frac{\Lambda^4}{\dot\phi^{3/2}\,H^{1/2}\,f_\phi^{1/2}} \, .
\label{gaussamps}
\end{equation}
For $f\sim 0.1\,m_{pl}$ and $\Lambda\ll 0.3\,M_{gut}$, as required by the slow-roll condition~(\ref{lambdabd2}), this correction is much smaller than ${\cal {O}}(1)$, as required by observations. 

Our derivation assumed that this shift led to a small change in $f_{NL}$ compared to $f_{NL, res}$.  This will be true if
\be
	f_{NL, nonres} \frac{\Delta {\cal P}}{{\cal P}} \ll f_{NL, res} \, .
\ee

Let us estimate the size of these effects.  Using $\dot{\phi}\sim H \mpl \sqrt{\eps}$, and approximating $H, \eps$ by their values for the purely quadratic potential, we can write (\ref{nongauss})\ as
\be
	f_{NL, res} \sim \Lambda^4 \sqrt{\frac{\sqrt{\eps} \mpl}{H^4 f_\phi^5}} \, .
\ee
As an example, let us choose $f_\phi\sim 0.1\mpl$, $\Lambda \sim 0.3\mpl$, where slow roll is starting to break down.  In this case $f_{NL,res} \sim 34$.
Let us note the strong dependence of $f_{NL,res}$ on the scale $\Lambda$, with a change of a factor of $3$ or so in $\Lambda$ inducing a change of a factor $\sim 100$ in $f_{NL,res}$. Since the currents bounds require $f_{NL}\la 100$ whereas it will be difficult to measure $f_{NL}\la 1$, such a change is sufficient to go from the regime already ruled out to a regime where $f_{NL}$ will be undetectable.

Finally, the contribution to $f_{NL}$ of the enhanced two-point function will be quite small. Using $f_{nonres} \sim 10^{-2}$, and using the tree-level results in \S5.1\ for  $H,\dot{\phi}$, we find
\be
	\frac{\delta{\cal P}}{{\cal P}} f_{nonres} \sim 4\times 10^{7}\left(\frac{\Lambda}{\mpl}\right)^4 \sqrt{\frac{\mpl}{f_\phi}} \, .
\ee
This is less than ${\cal O}(1)$ if $\Lambda \leq \left(f_\phi/m_{pl}\right)^{1/8} M_{gut}$. If the resonant contribution to the 3-point function is large, we can reasonable expect it to dominate $f_{NL}$.

\subsection{Nonperturbative transitions during inflation}

The potential energy of the inflaton in our model has multiple branches, depending on the amount of four-form flux at $\phi = 0$.  These branches can be labeled by the location $q = n e$ of their minima. Membranes charged under the four-form can nucleate and change branches. More generally, however, the other parameters  of the model depend on moduli and will also be landscape variables.  It is reasonable to expect that there will be multiple possible values for $\mu$ and that this coupling can also change via nonperturbative tunneling or bubble nucleation processes. For inflation to work, these processes must be rare to nonexistent during inflation.  On the other hand, if a tunneling event occurs early during the observable epoch of inflation, it may lead to an interesting observable signature.  We will sketch these statement here and leave a detailed exploration for future work.

Recall that in Minkowski space, in the thin wall approximation, the leading order bubble nucleation rate is (\ref{eq:decayrate}).  The bubble size is, explicitly, $r_0 = 3 \sigma/|\Delta V|$.  These equations can be easily derived (including the numerical factors) by balancing the energy of the domain wall against the energy difference inside and outside of the bubble.  The Minkowski approximation will be good if the bubble size is less than the Hubble scale; barring that, we should use the exact treatment with gravitational effects included \cite{Coleman:1980aw,Brown:1987dd,Brown:1988kg}.  

A general bubble will change both $n = q/e$ and $\mu$. Nonetheless, we will consider them independently.
Let us first consider bubbles which change $n$, assuming the thin wall approximation is good.  In this case, 
$\Delta V \sim \mu^2 \phi f_\phi = 2 V/N_w$, where $N_w$ is the number of windings.  We are assuming that the value of $\phi$ does not in fact change; only the branch of the monodromy potential changes.  This makes sense because the Compton wavelength $\mu^{-1} \sim 10 H^{-1}$; as we will find that $r_0 < H^{-1}$, $\mu^{-1} > r_0$.

For the argument of the exponential in (\ref{eq:decayrate}) to be of order unity, we must have
\be
	\sigma > \left(\frac{2 V^3}{27\pi^2 N_w^3}\right)^{1/4} \, .
\ee
For $f_\phi \sim 0.1\mpl, N_w \sim 100$, this means that $\sigma \sim (0.2 \times V^3)^{1/4}$, which means that the associated energy scale $\sigma^{1/3}$ is a bit below the GUT scale. Since the probability scales as $(\sigma^{1/3})^{12}$ it is not difficult to keep bubble nucleation suppressed. Note also that $\ln P$ increases as $1/V^3$ as $\phi$ decreases.  Over a single efold  $\Delta \phi^2 \sim 2\mpl^2/3$, and the action changes multiplicatively by a factor of $2.8$. The probability of bubble nucleation will fall off quickly.

Finally,  $r_0 \sim 2 M_{gut}^{-1}$, so the bubble size is well below the Hubble scale, and the Minkowski approximation is correct. It is also well below the Compton wavelength of the inflaton, so we can treat the value of $\phi$ as constant.  If 
$\sigma^{1/3}$ increases by a factor of $\sim 4$, $r_0 \sim H^{-1}$ and we must take gravitational corrections into account. Note that the bubble size will increase as $1/\Delta V$ during inflation; in particularly, over an efolding the radius will increase by a factor of $1.4$.  The Hubble scale increases as the square root of this, so over time the relative size $r_0/L_H = r_0 H$ will increase and one may have to take gravitational effects into account. Nevertheless, the probability for nucleation will still be small if the parameters were chosen to render it small initially.

Next, let us consider bubbles which change $\mu$. These are more complicated due to (\ref{eq:axionrelation}), but here we will just outline some basic features.  Again, we will assume that the thin wall approximation is valid, and consider sub-Hubble bubbles. For changes in $\mu$, $\Delta V = 2 (\Delta\mu/\mu) V$.  Then
\be
	P \sim \exp\left[ -\frac{27\pi^2}{2} \frac{\sigma^4}{V^3} \left(\frac{\mu}{2 \Delta \mu}\right)^3\right]\ ,
\ee
and the bubble size is
\be
	r_0 \sim \frac{3 \sigma}{V} \left(\frac{\mu}{2\Delta\mu}\right)\ .
\ee
$\sigma$ and $\Delta\mu$ are completely model dependent.  If $\sigma \sim V^{3/4}$, and $\Delta \mu \ll \mu$, the action is heavily suppressed.  On the other hand, the radius can become large; if $\Delta \mu < 0.007\mu$, the bubbles will be Hubble scale. Again, the probability decreases and the bubble size increases over the course of inflation. Note again, however, that because $\sigma$ is an energy scale cubed, a small change in the energy scale can still change the probability $P$ substantially.  For example, let $\Delta\mu \sim 0.1\mu$. Then the prefactor in the exponential is of order unity if 
$\sigma^{1/3} \sim 0.8 V^{1/4}$, and the probability is controlled by $(\mu/\Delta \mu)^3$ which can easily be large, and the nucleation probability exponentially small.

In both cases, the nucleation probability at the onset of inflation is highly sensitive to details of the models and could be appreciable at the onset of inflation, It will then die off exponentially in the number of efoldings. There are several interesting phenomenological possibilities in this case.  First, if the bubble nucleation happens just before the observable epoch of inflation, the interior of the bubble geometry will describe inflation with open spatial slices.  The observed universe can have a small negative spatial curvature which is potentially observable in future experiments \cite{Freivogel:2005vv}.  In a completely different scenario, one may imagine a setup whereby the membrane emission of either kind dominates over slow roll, leading to a form of chain inflation \cite{Freese:2006fk}. In such a case one has to carefully consider the quantum effects.  Clearly the subject of bubble nucleation in this scenario deserves further attention, and may offer a rich arena to study additional signatures of inflation on the sky.

\section{Conclusion}

The upshot of our paper is that high-scale inflation driven by a quadratic potential, realized via axion monodromy, is viable from  the standpoint of effective field theory.  The strongest constraints are the potential appearance of light states at large values of $F$, and the stiff constraints $M > H$ on moduli coupling to the inflaton sector. This latter constraint is nontrivial to satisfy in known string compactifications.

One reason models with observable primordial gravity waves are theoretically interesting is that they would require inflationary energy densities of order $M_{gut} \sim 2\times 10^{16}\ GeV$.  This implies that physics at the GUT scale could be the key to inflation.
In string and M-theory models consistent with grand unification, the fundamental string and/or 10/11-dimensional Planck scales are close to the GUT scale, often within a factor of 10 or so (see for example \cite{Kaloper:2002uj}, and references therein, for a discussion.) The effects we have discussed here -- indirect corrections to the four-form Lagrangian, direct sinusoidal corrections,  bubble nucleation -- are sensitive to physics at these scales.  It appears that a scenario with ${\cal O}(60)$ efoldings of inflation is close to the edge of viability.  The corrections to the basic quadratic inflation scenario -- such as resonant non-Gaussianity or bubble-induced features in the CMB spectrum  -- could be observable as a result.  

This intuition must be probed further.  If it holds up, it leads us to an interesting situation. On the one hand, high-scale inflation may not be completely parametrically safe and may not yield an arbitrary number of efoldings.  On the other hand, if inflation is nonetheless viable, there is a reasonable chance of more observable features. In addition to the above effects, a short epoch of inflation allows for small spatial curvature, and allows for some sensitivity to initial conditions.  If the inflaton began in a state excited above the Bunch-Davies vacuum, while the excitation will inflate away after a few efoldings \cite{Kaloper:2002cs}, it could lead to additional observable effects \cite{Kaloper:2003nv}. The edge of respectability is often the most interesting place to be.


\section{Acknowledgements}

We are grateful to A. Adams, N. Arkani-Hamed, F. Denef, S. Dubovsky, R. Easther, R. Flauger, R. Gopakumar, M. Kleban, A. Linde, M. Luty, M. Marino, L. McAllister, M. Mulligan, A. Nicolis, M. Porrati, F. Quevedo, M. Roberts, S. Schafer-Nameki, E. Silverstein, P. Steinhardt, L. Susskind, A. Tomasiello, A. Westphal and J. Yokoyama for useful discussions. A.L. would like to thank the cosmology and HET groups at UC Davis, U Mass Amherst, and the Harish-Chandra Research Institute while this work was carried out. A.L. would like to profusely thank D. Jatkar for lifesaving OSX help above and beyond the call of duty. N.K. is grateful to the Aspen Center for Physics, to SITP, Stanford, to RESCEU and to IPMU, University of Tokyo, for kind hospitality during the course of this work. We would also like to thank the Michigan CTP, during the workshop on ``Effective Field Theory in Cosmology", the organizers of the ``Time and Matter" conference in Budva, Montenegro, and the ``Primordial Features and Non-Gaussianity" workshop at HRI for providing stimulating environments at various points during this project. N.K. is supported in part by the DOE Grant DE-FG03-91ER40674. A.L. is supported in part by DOE Grant DE-FG02-92ER40706. L.S. is supported in part by the NSF grant PHY - 0855119.

\vskip1cm

\bibliographystyle{utphys}
\bibliography{kls_refs}

\end{document}